\documentclass[aps,10pt,pra,floatfix,twocolumn,superscriptaddress]{revtex4-1}
\usepackage{color}
\usepackage{ulem}
\usepackage{graphicx}
\usepackage{subfigure,amsmath, amsthm, amssymb}
\usepackage[colorlinks,citecolor=red]{hyperref}
\usepackage{braket}
\usepackage{enumitem}

\newcommand{\be}{\begin{equation}}
\newcommand{\ee}{\end{equation}}
\newcommand{\bea}{\begin{eqnarray}}
\newcommand{\eea}{\end{eqnarray}}

\begin{document}

\title{Protecting quantum correlations in presence of generalised amplitude damping channel: the two-qubit case}

\author{Suchetana Goswami}
\email{suchetana.goswami@gmail.com}
\affiliation{S. N. Bose National Centre for Basic Sciences, Salt Lake, Kolkata 700106,  India}

\author{Sibasish Ghosh}
\email{sibasish@imsc.res.in}
\affiliation{Optics and Quantum Information Group, The Institute of Mathematical Sciences, C. I. T. Campus, Taramani, Chennai 600113, India}
\affiliation{Homi Bhabha National Institute, Training School Complex, Anushakti Nagar, Mumbai 400094, India}

\author{A. S. Majumdar}
\email{archan@bose.res.in}
\affiliation{S. N. Bose National Centre for Basic Sciences, Salt Lake, Kolkata 700106,  India}

\date{\today}

\begin{abstract}

Any kind of quantum resource useful in different information processing tasks is vulnerable to several types of environmental noise. Here we study the behaviour of quantum correlations such as entanglement and steering in two-qubit systems under the application of the generalised amplitude damping channel and propose two protocols towards preserving them under this type of noise. First, we employ the technique of weak measurement and reversal for the purpose of preservation of correlations. We then show how the evolution under the channel action can be seen as an unitary process. We use the technique of weak measurement and most general form of selective positive operator valued measure (POVM) to achieve preservation of correlations for a significantly large range of parameter values.

\end{abstract}

\maketitle

\section{Introduction}

Non-local features of quantum correlations enable us to perform various information processing tasks, such as quantum teleportation \cite{BBCJPW_93}, super dense coding \cite{BW_92}, quantum error correction \cite{S_95}, device-independent quantum key distribution \cite{BB_84, E_91, BCWSW_12}. While practically implementing such tasks, there is always an interaction of the concerned quantum system with some noisy environment. Such interactions diminish the quantum correlations present in the state under consideration, in general. So it is one of the most important jobs in any information processing task to preserve the quantum correlations in presence of noisy environment, at least by some amount. This necessitates formulation of protocols for controlling decoherence of a system. 

There are various well-known forms of decoherence modelled by qubit channels such as the depolarising channel, dephasing channel, amplitude damping channel (ADC), generalised amplitude damping channel (GADC), and so on \cite{BFMP_00, SN_96, SB_08, F_04, NC_02}. It  is not possible to enhance quantum correlation in two-qubit system by unital operations, whereas for some initial states it might be possible to enhance or generate quantum correlation such as discord when the system passes through some non-unital channel \cite{SKB_11}. For example, interaction through amplitude damping channel can enhance the teleportation fidelity for a particular class of two-qubit entangled state \cite{BHHH_00, B_02}. On the other hand, using the technique of weak measurement, one can protect loss of entanglement \cite{KU_99, KJ_06, KC_09, KL_12, HMK_20}, and improve the fidelity of teleportation \cite{PM_13} as well as the secret key rate for certain quantum key distribution protocols \cite{DGPM_17}, while the interaction is taking place through the ADC. 

There exist other ways to protect quantum correlations from environmental noise, such as by employing quantum Zeno effect \cite{MS_77,FLP_04}, frequent unitary interruptions (bang-bang pulse) \cite{VL_98, Z_99, BL_02, BL_03}, strong continuous coupling \cite{P_80, S_98, FTPNTL_05}, etc. In the present work, we focus on the problem of preservation of quantum correlations under the decoherence arising from the action of the generalised amplitude damping channel. Note that all of the aforesaid decoherence control process are dynamical in nature: one needs to follow the dynamics of the system in order to implement each such control process. On the other hand, the environment action, considered in the present work is of static nature and hence the aforesaid decoherence controlling mechanisms will not work, in general, in the case considered here.

In quantum information theory, there exists mainly three types of nonlocal correlations for multipartite systems, namely entanglemet \cite{EPR_35}, steering \cite{S_35, R_89, WJD_07, JWD_07, SNC_14, GA_15} and Bell-nonlocality \cite{B_64, CHSH_69, W_89}. In 1935, the concept of entanglement was first introduced in context of the famous EPR paradox \cite{EPR_35}. In the same year, this paradox was revisited by Schr\"{o}dinger and was interpreted with the introduction of another stricter form of quantum correlation for pure states, termed as steerability \cite{S_35}. Much later, this concept was generalised for mixed states \cite{WJD_07, JWD_07}. Meanwhile, John Bell introduced the idea of Bell-nonlocality \cite{B_64} which is the strongest form of quantum correlation known so far.

Here, we confine our studies within the first two types of quantum correlations, namely, entanglement and steering. Each of these two correlations decreases under the action of generalised amplitude damping channel acting on one of the qubits, which is a non-unital channel. Now, for the purpose of preservation of non-local correlation, we start with pure (maximally and non-maximally) entangled states, and first employ the technique of weak measurement which has been used in case of the standard ADC \cite{DGPM_17}. 

In the case of weak measurement, as first proposed in \cite{AAV_88}, the interaction between the system and the apparatus is taken to be very weak, along with two processes termed as pre-selection and post-selection measurement\cite{DMMJAB_14}. To use weak measurement as a procedure for preservation of correlations in a quantum state, one has to do the weak measurement, and the reverse weak measurement to be followed at the end of the protocol. The procedure of weak measurement has been used in a huge number of protocols to study different interesting phenomena in quantum theory, such as spin Hall effect \cite{HK_08}, wave particle duality using cavity-QED experiments \cite{W_02}, superluminal propagation of light \cite{AMCPH_04,BSWLG_04}, direct measurement of the quantum wave function \cite{LSPSB_11}, measurement of ultrasmall time delays of light \cite{BS_10}, observing Bohmian trajectories of photons \cite{KBRSMSS_11, GMGS_01} and also for detection of entanglement with minimal resources \cite{GCGM_18}. 

Next, to make the technique of preservation of non-local correlations more general for environmental noise, we find the unitary dilation corresponding to the completely positive trace-preserving evolution of the GADC, starting from its known Kraus representation \cite{BP_02}. After finding the inverse of this unitary, we construct the most general form of operator-sum representation (Kraus representation) of an approximate inverse map, which is not unique. Employing these Kraus operators individually as the elements of a POVM, we show that it is possible to preserve the correlations of the initial state up to certain extents for a broad range of state parameter and damping coefficients. 

It should be noted that in all cases here one has to employ selective POVM, as non-selective POVM corresponds to a unitary evolution, and under local unitary it is not possible to generate or enhance any kind of quantum correlation \cite{VP_98}. Although our method (to be described in Sec. (\ref{sec_unitary}) below) may appear to be quite specific towards tackling the noise of GADC, nevertheless, as a method, it has a general appeal in the sense that it can, in principle, be applicable to  any noise model - provided we have the prior information about the noise model. (It may be note here that  the aforesaid process of addressing decoherence via unitary dilation does not take into account, in general, any driving mechanism, to act on the system using apparatus having support outside of the ancilla Hilbert space. In this sence, our aforesaid approach is a restricted one.) Any physical quantum operation is guided by a completely positive trace preserving (CPTP) map which in turn can be expressed as an unitary dilation. The applicability of unitary dilation is a well discussed topic in literature \cite{BDPS_03, FEGVH_11, BDS_18}. Also the implementation of selective POVM is nothing but doing some probabilistic measurements and can also be physically realized \cite{ZB_05}. 

Our paper is arranged in the following way. After the introduction, we first discuss the preliminary definitions related to our protocol and the technical tools that we will  use in the rest of the analysis, in Sec. (\ref{background}). In Sec. (\ref{sec_weak}), we discuss our first approach for the preservation of quantum correlations using the technique of weak measurement. In the following Sec. (\ref{sec_unitary}), we propose another  more general protocol for the preservation of quantum correlations considering GADC as a unitary evolution. In  Sec. (\ref{concl}), we summarize our work, and present a brief discussion about some future directions.

\section{Background}
\label{background}

Any quantum state that can be written as a convex mixture of product states of its own subsystems, is termed as a separable state. Now any state that can not be expressed in this form, is an entangled state, i.e. for a bipartite entangled state $\rho_{AB}^{E}$, one can write,
\begin{eqnarray}
\rho_{AB}^{E} \neq \sum_{i} p_{i} \rho_{A}^i \otimes \rho_{B}^i.
\label{entstate}
\end{eqnarray}
for any $\rho_{A}^i$ and $\rho_{B}^i$, the states corresponding to the subsystems $A$ and $B$ respectively.  There are various ways to quantify entanglement in a given biparite state \cite{GT_09}. For the purpose of this problem, we consider concurrence as the measure of entanglement \cite{CKW_00}. Concurrence of a given two-qubit state $\rho_{AB}$ is defined as
\begin{eqnarray}
C(\rho_{AB})=max\lbrace(\sqrt{\lambda_{1}}-\sqrt{\lambda_{2}}-\sqrt{\lambda_{3}}-\sqrt{\lambda_{4}}),0\rbrace.
\label{conc}
\end{eqnarray}
with, $\lambda_{1}$, $\lambda_{2}$, $\lambda_{3}$, $\lambda_{4}$ being the eigenvalues of the matrix $\rho_{f}$ in the descending order, with $\rho_{f}=\rho_{AB} . \tilde{\rho}_{AB}$ ($\tilde{\rho}_{AB}= (\sigma_{y}\otimes\sigma_{y}).{\rho_{AB}}^*.(\sigma_{y}\otimes\sigma_{y})$, $\sigma_{y}$ being the Pauli Y-matrix). 

Now, let us briefly discuss about another quantum correlation, termed as quantum steering which is a stronger form of quantum correlation than entanglement. In this scenario, one side (say, $A$) is untrusted i.e., the dimension of the subsystem $A$ is completely uncharacterised. If $A$ can convince $B$ by local operation and classical communication (LOCC), that they are sharing an entangled state, then the state shared between them is said to be a steerable state. Here we consider the steering scenario, in which $A$ performs two black-box dichotomic observables (i.e. they can be any two dichotomic observables in any dimension) $A_0$ and $A_1$ and $B$ performs two qubit-measurements in mutually unbiased bases, given by $B_0~(=\sigma_{z})$ and $B_1~(=\sigma_x)$. Then the necessary and sufficient condition for quantum steering \cite{CFFW_15} can be written in terms of an inequality (also termed as ACHSH inequality for steering) as stated below. 
\begin{eqnarray}
&\sqrt{\langle (A_{0} + A_{1}) B_{0} \rangle^2 + \langle (A_{0} + A_{1}) B_{1} \rangle^2 } \nonumber \\
&+\sqrt{\langle (A_{0} - A_{1}) B_{0} \rangle^2 + \langle (A_{0} - A_{1}) B_{1} \rangle^2 } \leq 2, 
\label{achsh}
\end{eqnarray}
with, $\langle A_x B_y \rangle =  \sum_{a,b} (-1)^{a\oplus b} p(ab|xy)$.  Here, $a$ and $b$ are the possible outcomes ($\in \{ 0,1 \}$) of the dichotomic measurements by $A_x$ and $B_y$ respectively. We will be considering the inequality given in Eq.(\ref{achsh}) for the certification of steerability in this paper. Note that the violation of this inequality is necessary and sufficient to certify steering in a two-qubit system.

\begin{figure*}[ht]
\resizebox{14cm}{8cm}{\includegraphics{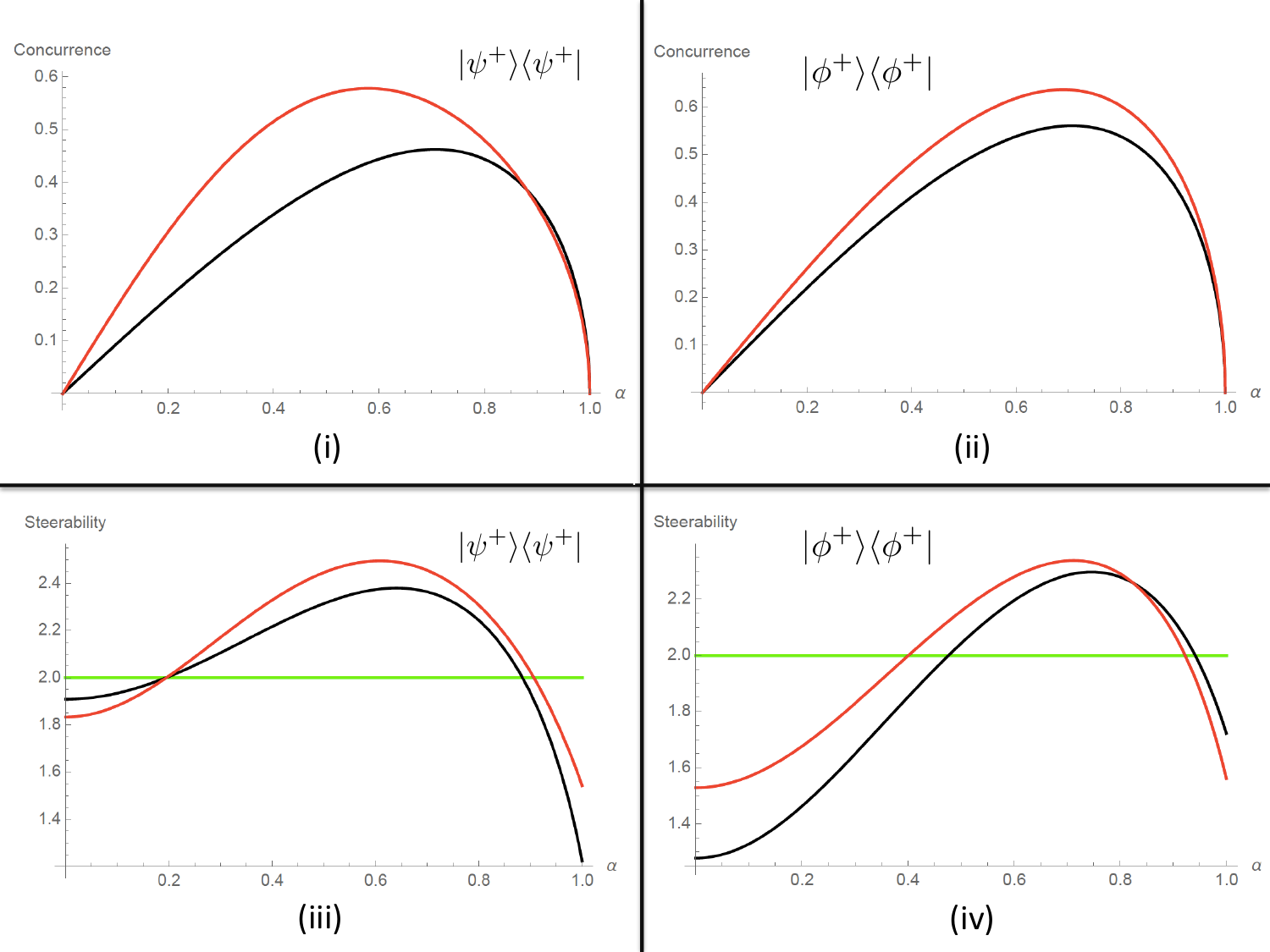}}
\caption{\footnotesize (Colour online) (i) Comparison of concurrence while the initial state is taken to be the antiparallel state, $\ket{\psi^{+}}\bra{\psi^{+}}$ ($\nu= 0.958$ , $\eta= 0.324$), (ii) Comparison of concurrence while the initial state is taken to be the parallel state, $\ket{\phi^{+}}\bra{\phi^{+}}$ ($\nu= 0.892$ , $\eta= 0.447$), (iii) Comparison of steerability while the initial state is taken to be the antiparallel state, $\ket{\psi^{+}}\bra{\psi^{+}}$ ($\nu= 0.895$ , $\eta= 0.660$), (iv) Comparison of steerability while the initial state is taken to be the parallel state, $\ket{\phi^{+}}\bra{\phi^{+}}$ ($\nu= 0.787$ , $\eta= 0.749$). In all the plots, Red curves denote the corresponding function with the weak measurement and Black curves denote the same without weak measurement, where Green lines in the plots (iii) and (iv) denote the limit of the violation of the ACHSH inequality. (For individual plots, $\nu$, $\eta$ are mentioned.)}
\label{plot_weak}
\end{figure*}

In this paper, we study the behaviour of the aforesaid two quantum correlations (entanglement and steering) under the application of local environmental noise expressed in the form of generalised amplitude damping. GADC can be obtained by solving the optical master equation in presence of a squeezed thermal bath and this channel describes the effect of environmental dissipation in a finite temperature bath. The Kraus operators of the corresponding channel are given by,
\begin{eqnarray}
K_{1}=\sqrt{\nu}\begin{bmatrix}
1&0\\
0&\sqrt{\eta}
\end{bmatrix};~
K_{2}=\sqrt{\nu}\begin{bmatrix}
0&\sqrt{1-\eta}\\
0&0
\end{bmatrix} \nonumber \\
K_{3}=\sqrt{1-\nu}\begin{bmatrix}
\sqrt{\eta}&0\\
0&1
\end{bmatrix};~
K_{4}=\sqrt{1-\nu}\begin{bmatrix}
0&0\\
\sqrt{1-\eta}&0
\end{bmatrix}.
\label{Kraus_GADC}
\end{eqnarray}
where, $\nu \in [0,1]$ is related to the temperature of the heat bath and $\eta \in [0,1]$ is the parameter representing the rate of dissipation due to the heat bath action. Note that, for $\nu=1$, the Kraus representation in Eq.(\ref{Kraus_GADC}) traces down to that of an ADC, for which the environment is assumed to be at zero temperature. The GADC map $\Lambda$ acts here on one of the qubits of a two-qubit state 
\begin{eqnarray}
\rho_{AB}: \Lambda(\rho_{AB}) = \sum_{i=1}^{4} (\openone_{A} \otimes K_{i}) \rho_{AB} (\openone_{A} \otimes K_{i}^\dagger).
\label{GADC_evolution}
\end{eqnarray}

In previous works it has been shown that it is possible to subdue the effect of environmental interaction through ADC by the application of the technique of weak measurement and its reversal \cite{KU_99, KJ_06, KC_09, KL_12, PM_13, DGPM_17}. In the similar way, in this paper, we first study the effect of weak measurement technique when the environmental interaction is taking place through GADC. Here, before the environmental interaction takes place with the particle in consideration (say, $B$) of the bipartite system $AB$, weak measurement (WM) with a strength $w$ is performed on the same. The detector detects the system with probability $w$ if and only if the state of $B$ is in $\ket{1} (= \begin{bmatrix}
0\\
1
\end{bmatrix})$ and hence the measurement operator $W_{1}$ corresponding to this scenario is given as,
\begin{eqnarray}
W_{1}= \sqrt{w} \ket{1}\bra{1} = \begin{bmatrix}
0&0\\
0& \sqrt{w}
\end{bmatrix}.
\label{Kraus_weak_1}
\end{eqnarray}
Note that, the matrix in Eq. (\ref{Kraus_weak_1}) is singular. Hence this is not effective to implement for the reverse weak measurement. So, for the purpose of weak measurement, we consider the measurement operator corresponding to the scenario when the system is not detected by the measuring apparatus. The measurement operator $W_{0}$ corresponding to this situation can be evaluated by using the relation, $W_{1}^{\dagger} W_{1}+ W_{0}^{\dagger} W_{0} = \openone$. Hence,
\begin{eqnarray}
W_{0}= \ket{0}\bra{0}+ \sqrt{1-w} \ket{1}\bra{1}=\begin{bmatrix}
1&0\\
0& \sqrt{1-w}
\end{bmatrix}.
\label{Kraus_weak_0}
\end{eqnarray}

As the matrix in Eq. (\ref{Kraus_weak_0}) is a reversible one, application of the inverse of it leads back the system to its original state. According to the our protocol, after the weak measurement is done, the particle in consideration interacts with the environment through the GADC and lastly reverse weak measurement (RWM) is done on the same. The Kraus opearator corresponding to the reverse weak measurement is gievn below.
\begin{eqnarray}
R_{0}=\begin{bmatrix}
\sqrt{1-r} &0\\
0& 1
\end{bmatrix}.
\label{Kraus_revweak_0}
\end{eqnarray}
where, $r$ is the strength of the reverse weak measurement (we consider it different from the weak measurement strength $w$ to make sure that there is a freedom of choice for different efficiencies of weak and reverse weak measurement). 

After the implementation of the technique of weak measurement, now we propose another, our more general approach for the purpose of preservation of steering and entanglement of the bipartite state while interacting with environment through GADC. Here we consider the unitary dilation corresponding to the completely positive trace preserving (CPTP) map described by the GADC. The unitary dilation corresponding to the Kraus operator representation given in Eq.(\ref{Kraus_GADC}), is not unique and can be obtained considering a two-qubit ancilla for the action of GADC on one side of the two-qubit system. The action of this unitary (say, $U_{SB}$) on the initial state of system ($B$) plus ancilla ($S$) gives the state after the environmental interaction has taken place through GADC. In the next step, we find the inverse of this unitary ($U_{SB}^{-1}$) from which one can find the corresponding Kraus operator representation of the evolution. These individual Kraus operators can be considered as elements of the most general POVM \cite{note}. Employing the selective POVM obtained corresponding to the inverse map, either before or after the action of the GADC, we study the concurrence and the steerability of the final state.

\section{Employing the technique of weak measurement}
\label{sec_weak}

As described in the previous section, here we consider that only one side of the bipartite system is interacting with the environment. We study two cases described in the flowchart given below.\\
\begin{itemize}
\item[$\rho_{AB}$]  $\xrightarrow{GADC (B)} \rho_{AB}^{\prime} =  \sum_{i=1}^{4} (\openone \otimes K_{i}) \rho_{AB} (\openone \otimes {K_{i}}^\dagger)$.
\item[$\rho_{AB}$]  $\xrightarrow{WM (B)} \rho_{AB}^{w} = \frac{(\openone \otimes W_{0}) \rho_{AB} (\openone \otimes {W_{0}}^\dagger)}{Tr[(\openone \otimes W_{0}) \rho_{AB} (\openone \otimes {W_{0}}^\dagger)]}$ \\$ \xrightarrow{GADC (B)} \rho_{AB}^{d} = \sum_{i=1}^{4} (\openone \otimes K_{i}) \rho_{AB}^{w} (\openone \otimes {K_{i}}^\dagger)$ \\ $ \xrightarrow{RWM (B)} \rho_{AB}^{\prime \prime}= \frac{(\openone \otimes R_{0}) \rho_{AB}^{d} (\openone \otimes {R_{0}}^\dagger)}{Tr[(\openone \otimes R_{0}) \rho_{AB}^{d} (\openone \otimes {R_{0}}^\dagger)]}$. \\
\end{itemize}
where $K_{i}$'s, $W_{0}$ and $R_{0}$ are given by Eq. (\ref{Kraus_GADC}), (\ref{Kraus_weak_0}) and (\ref{Kraus_revweak_0}) respectively. Note that comparison of steerability and concurrence between the states $\rho_{AB}^{\prime}$ and $\rho_{AB}^{\prime \prime}$ gives the idea about the fact whether the technique of weak measurement is useful for the preservation of quantum correlation. In the whole paper, we consider either pure anti-parallel entangled state ($\ket{\psi^{\pm}}\bra{\psi^{\pm}}$) or pure parallel entangled state ($\ket{\phi^{\pm}}\bra{\phi^{\pm}}$) 
in computational basis as the initial state ($\rho_{AB}$) for all the protocols, where,
\begin{eqnarray}
\ket{\psi^{\pm}}=\alpha \ket{01} \pm \beta \ket{10}, 
\label{initial_antiparallel} \\
\ket{\phi^{\pm}}=\alpha \ket{00} \pm \beta \ket{11}.
\label{initial_parallel}
\end{eqnarray}
In the above cases one must have $\alpha^2+\beta^2=1$, to fulfil the demand of normalisation. 
We compute the concurrence as well as the magnitude of violation of the analog CHSH inequality as a quantifier for steering.
It is evident from Fig.(\ref{plot_weak}), the correlations (entanglement and steering) show a certain amount of improvement for a section of pure states (for some values of state parameter $\alpha$) under the application of weak measurement technique. Note that the plots in Fig.(\ref{plot_weak}) are corresponding to a particular set of values of the GADC parameters, $\nu$ and $\eta$. It can be seen that for some other set of values for the channel parameter $\nu$ and $\eta$, it is possible to preserve both the correlations under the application of weak measurement corresponding to different weak measurement strength $w$ and reverse weak measurement strength $r$. But, the range of channel parameters for this technique showing any improvement is quite small.

\begin{figure}[ht]
\resizebox{8cm}{5cm}{\includegraphics{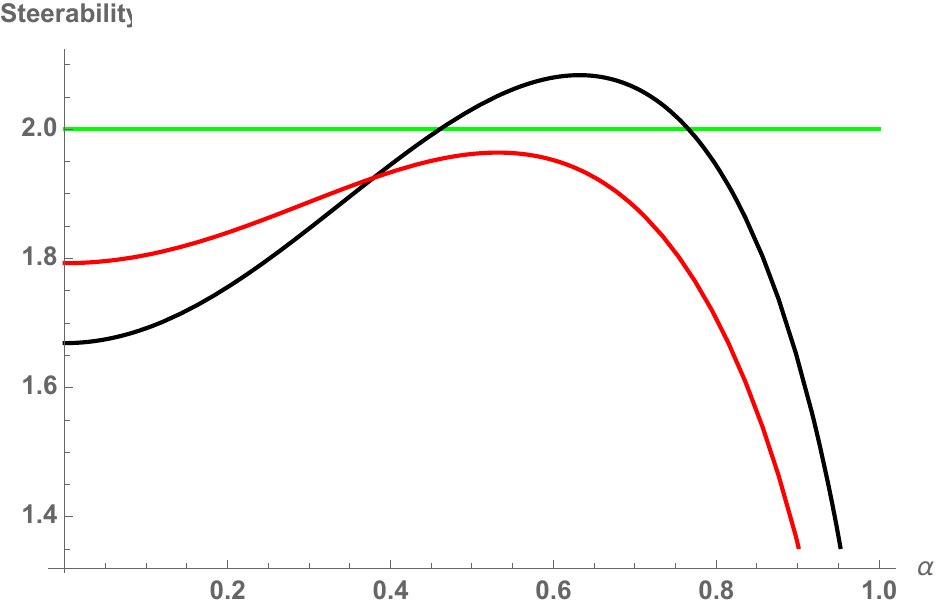}}
\caption{\footnotesize (Colour online) The plot is the same as in Fig (1 (iv)) but corresponding to a different set of channel parameter i.e. $\nu=0.234$ and $\eta=0.646$. The colours have their usual meaning.}
\label{particular_weak}
\end{figure}

It may be noted that for a large number of values (though not always for a continuous range) of $\nu$ and $\eta$ this technique is unable to preserve the correlations. In fact, this technique  has the capability of functioning for different weak and reverse weak measurement strengths and giving improvement for a rather restrictive set of parameter values of channel and state parameter (one of which is shown in the Fig (\ref{plot_weak})). On the contrary, for other values of the channel parameter, the state remains un-steerable even after employing weak measurement technique. For example, for $\nu=0.234$ and $\eta=0.646$ this fact can be observed from Fig (\ref{particular_weak}). So in the next section we propose a more general approach for the preservation of quantum correlation using the selective POVM. This approach is general in the sense that it deals with the unitary dilation of the GADC, which is not unique. This gives us the freedom to choose the suitable POVM that pre
 serves the correlation maximally. 

\section{Preserving the correlation considering a unitary dilation of GADC}
\label{sec_unitary}

It is a well-known fact that any quantum channel that corresponds to a physical process can be seen as a CPTP evolution. In this section we try to obtain the unitary dilation of the CPTP evolution corresponding to the GADC and use the same for the purpose of preservation of entanglement and steerability.

\subsection{Motivation behind the approach}
\label{Mot_app}

Every CPTP map ($\mathcal{N}$) can be realized via unitary dilation, followed by tracing out the ancillary part: $\mathcal{N} (\rho_S) = Tr_{B} [U_{SB} (\rho_S \otimes \sigma_B^{(0)}) {U_{SB}}^\dagger]$ with a fixed state $\sigma_B^{(0)}$ of the ancilla and any state $\rho_S$ of the system $S$. Such a unitary is not unique even if we allow the dimension of the ancilla to be square of the minimum number of Kraus operators needed to realize $\mathcal{N}$. Our aim is to choose such an $U_{SB}$ (as well as $\sigma_B^{(0)}$) such that: \\
i) $U_{SB}$ becomes $\openone_S \otimes \openone_B$ whenever $\mathcal{N}$ becomes the identity channel.\\
ii) $U_{SB} (\rho_S \otimes \sigma_B^{(0)}) {U_{SB}}^\dagger$ is close to a product state of the form $Tr_{B} [U_{SB} (\rho_S \otimes \sigma_B^{(0)}) {U_{SB}}^\dagger] \otimes \sigma_B^{(1)}$ where $\sigma_B^{(1)}$ is fixed state of the ancilla. 

Condition (ii) guarantees that $U_{SB}^{-1} [Tr_{B} \lbrace U_{SB} (\rho_S \otimes \sigma_B^{(0)}) {U_{SB}}^\dagger \rbrace \otimes \sigma_B^{(1)}] ({U_{SB}^{-1}})^\dagger$ is close to the initial state $\rho_S \otimes \sigma_A^{(0)}$. And hence, the reduced state of $S$, after the action of $U_{SB}^{-1}$ would become close to $\rho_S$. Moreover $\rho_S^{\prime} \rightarrow Tr_{B}[U_{SB}^{-1} (\rho_S^{\prime} \otimes \sigma_B^{(1)}) ({U_{SB}^{-1}})^\dagger]$ is a CPTP map on $S$. This is the motivation behind adopting the aforesaid approach.

The pertinent question that one needs to answer is whether it is at all possible to satisfy condition (ii) (condition (i) can always be satisfied). Though there exists no general proof of this as such, we  consider here (in the next section) two types of such unitaries $U_{SB}$'s, and validity of condition (ii) may be tested for these two choices.

\subsection{To find the unitary dilation and its inverse}

Let us start by considering a general CPTP map given as $\mathcal{N}: \mathcal{B}(\mathcal{H}_S) \rightarrow \mathcal{B}(\mathcal{H}_S)$, with $\mathcal{H}_S$ being the $d$ dimensional Hilbert space associated with a given quantum system $S$ and $\mathcal{B}(\mathcal{H}_s)$ represents the set of all bounded linear operators, $\mathcal{A}: \mathcal{H}_S \rightarrow \mathcal{H}_S$. It is obvious that $\mathcal{N}$ has an operator-sum representation (or, Kraus representation) expressed as, $\mathcal{N}(\mathcal{A})=\sum_{i=1}^{M} L_i \mathcal{A} {L_i}^\dagger$ ($M$ being a finite positive integer) with the Kraus operators $L_i$ satisfying the condition, $\sum_{i=1}^{M} {L_i}^\dagger L_i = \openone_d$. Now, let us consider an ancilla system $B$ whose associated Hilbert space $\mathcal{H}_B$ has dimension $M$. Let, $\lbrace \ket{i}_S \rbrace_{i=1}^d$ be an orthonormal basis (ONB) for $\mathcal{H}_S$ while $\lbrace \ket{i}_B \rbrace_{i=1}^M$ be an ONB for $\mathcal{H}_B$. Our aim i
 s to find a $dM \times dM$ unitary matrix $U_{SB}$ which corresponds to the map $\mathcal{N}$ such that,
\begin{eqnarray}
L_i= {}_{B}\bra{i} U_{SB} \ket{1}_B, \forall i=1,2,...,M.
\label{KraustoUni1}
\end{eqnarray}
Alternatively, for every $\mathcal{A} \in \mathcal{B}(\mathcal{H}_S)$ (thus, $\mathcal{A}$ can also be a density matrix of $S$), one can write,
\begin{eqnarray}
\mathcal{N}(\mathcal{A})=Tr_{B} [U_{SB} (\mathcal{A} \otimes \ket{1}_{B} \bra{1}) {U_{SB}}^\dagger].
\label{KraustoUni2}
\end{eqnarray}
Note that, for a given CPTP map, its unitary dilation given by the matrix $U_{SB}$ is not unique. The $(\alpha i, \beta 1)$-entry of the matrix can be obtained in the following way,
\begin{eqnarray}
u_{\alpha i, \beta 1}&\equiv  ({}_S\bra{\alpha} \otimes {}_B\bra{i}) U_{SB} (\ket{\beta}_S \otimes \ket{1}_B) \nonumber\\
& ={}_S \bra{\alpha} L_{i} \ket{\beta}_S,~ \forall \alpha,\beta =1,2,...,d.
\label{uni_element}
\end{eqnarray}
For a given set of Kraus operators, $L_1$, $L_2$,..., $L_M$, with the help of Eq. (\ref{uni_element}), it is possible to obtain information about $d$ column vectors of the unitary matrix $U_{SB}$ (with respect to the joint ONB $\lbrace \ket{\alpha} \otimes \ket{i} \big{\vert} \alpha=1,2,...,d; i=1,2,...,M \rbrace $) and they are given below. 
\begin{eqnarray}
\overrightarrow{u_{11}}=  (u_{11,11}, & u_{12,11},..., u_{1M,11}, u_{21,11}, u_{22,11},..., \nonumber\\
& u_{2M,11},..., u_{d1,11}, u_{d2,11},..., u_{dM,11})^T, \nonumber\\
\overrightarrow{u_{21}}=  (u_{11,21}, & u_{12,21},..., u_{1M,21}, u_{21,21}, u_{22,21},..., \nonumber\\
& u_{2M,21},..., u_{d1,21}, u_{d2,21},..., u_{dM,21})^T, \nonumber\\
. .\nonumber\\
\overrightarrow{u_{d1}}=  (u_{11,d1}, & u_{12,d1},..., u_{1M,d1}, u_{21,d1}, u_{22,d1},..., \nonumber\\
& u_{2M,d1},..., u_{d1,d1}, u_{d2,d1},..., u_{dM,d1})^T
\label{d_column}
\end{eqnarray}
Thus, we obtain an incomplete ONB $\lbrace \overrightarrow{u_{11}}, \overrightarrow{u_{21}},..., \overrightarrow{u_{d1}} \rbrace$ of the $dM$ dimensional Hilbert space $\mathcal{H}_S \otimes \mathcal{H}_B$. At this point, we use the method of basis extension to extend this incomplete ONB given in Eq. (\ref{d_column}) to form the complete ONB $\lbrace \overrightarrow{u_{11}}, \overrightarrow{u_{12}},..., \overrightarrow{u_{1M}}, \overrightarrow{u_{21}}, \overrightarrow{u_{22}},..., \overrightarrow{u_{2M}},..., \overrightarrow{u_{d1}}, \overrightarrow{u_{d2}},..., \overrightarrow{u_{dM}} \rbrace$ for $\mathcal{H}_S \otimes \mathcal{H}_B$ and eventually to construct the unitary matrix $U_{SB}$. The procedure of basis expansion is not unique, but it is restricted by two conditions: i) all the column vectors of $U_{SB}$ should be orthogonal to each other and ii) the individual columns must be normalised. Taking these two constraints into consideration one can construct different f
 orms of $U_{SB}$ starting from Eq. (\ref{d_column}), all of which represents the same CPTP map $\mathcal{N}$ that we have started with.

Now, let us consider the Kraus representation of GADC (expressed in the computational basis) given in Eq. (\ref{Kraus_GADC}) and the evolution of $\rho_{AB}$ is given by, Eq. (\ref{GADC_evolution}). Hence, in this particular scenario of environmental interaction through GADC, we have $M=4$ and $d=2$. Thus, in this case, the unitary matrix $U_{SB}$ must be of dimension $8 \times 8$. Following the technique mentioned above in Eq. (\ref{uni_element}), we find only two columns of the $8 \times 8$ unitary, which are $\overrightarrow{u_{11}}$ and $\overrightarrow{u_{21}}$,:
\begin{widetext}
\begin{eqnarray}
\overrightarrow{u_{11}} = (u_{11,11} \equiv {}_S \bra{0} K_{1} \ket{0}_S = \sqrt{\nu},~u_{12,11} \equiv {}_S \bra{0} K_{2} \ket{0}_S = 0,~u_{13,11} \equiv {}_S \bra{0} K_{3} \ket{0}_S = \sqrt{(1-\nu) \eta}, \nonumber \\
u_{14,11} \equiv {}_S \bra{0} K_{4} \ket{0}_S = 0,~u_{21,11} \equiv {}_S \bra{1} K_{1} \ket{0}_S = 0,~u_{22,11} \equiv {}_S \bra{1} K_{2} \ket{0}_S = 0, \nonumber\\
u_{23,11} \equiv {}_S \bra{1} K_{3} \ket{0}_S = 0,~u_{24,11} \equiv {}_S \bra{1} K_{4} \ket{0}_S = \sqrt{(1-\nu)(1-\eta)})^T; \\
\overrightarrow{u_{21}} = (u_{11,21} \equiv {}_S \bra{0} K_{1} \ket{1}_S = 0,~u_{12,21} \equiv {}_S \bra{0} K_{2} \ket{1}_S = \sqrt{\nu (1-\eta)},~u_{13,21} \equiv {}_S \bra{0} K_{3} \ket{1}_S = 0, \nonumber \\
u_{14,21} \equiv {}_S \bra{0} K_{4} \ket{1}_S = 0,~u_{21,21} \equiv {}_S \bra{1} K_{1} \ket{1}_S = \sqrt{\nu \eta},~u_{22,21} \equiv {}_S \bra{1} K_{2} \ket{1}_S = 0, \nonumber\\
u_{23,21} \equiv {}_S \bra{1} K_{3} \ket{1}_S = \sqrt{1-\nu},~u_{24,21} \equiv {}_S \bra{1} K_{4} \ket{1}_S = 0)^T.
\label{2_col_GADC}
\end{eqnarray}
\end{widetext}
Using the method of basis expansion, and by taking care of the constraints stated previously, we construct several unitary matrices representing the noisy channel (GADC) in consideration. Note that, from Eq. (\ref{Kraus_GADC}), for $\nu=\eta=1$, the channel should represent an identity operation. Keeping all these facts in mind, in this work, we concentrate on two separate unitary evolutions corresponding to the GADC which are given in terms of the $8 \times 8$ matrix $U_{SB}$ below.
\begin{widetext}
\begin{eqnarray}
U_{SB}^{(1)} &=& \tiny{\left(
\begin{array}{cccccccc}
 \sqrt{\nu } & 0 & -\frac{\sqrt{(1-\nu ) \eta }}{\sqrt{1-(1-\nu ) (1-\eta )}} & 0 & 0 & 0 & 0 & -\frac{\sqrt{\nu  (1-\nu ) (1-\eta )}}{\sqrt{1-(1-\nu ) (1-\eta )}} \\
 0 & \sqrt{\eta } & 0 & 0 & \sqrt{\nu  (1-\eta )} & 0 & -\sqrt{(1-\nu ) (1-\eta )} & 0 \\
 \sqrt{(1-\nu ) \eta } & 0 & \frac{\sqrt{\nu }}{\sqrt{1-(1-\nu ) (1-\eta )}} & 0 & 0 & 0 & 0 & -\frac{(1-\nu ) \sqrt{\eta  (1-\eta )}}{\sqrt{1-(1-\nu ) (1-\eta )}} \\
 0 & 0 & 0 & \sqrt{\eta  (1-\nu )+\nu } & 0 & -\sqrt{(1-\nu ) (1-\eta )} & 0 & 0 \\
 0 & -\sqrt{1-\eta } & 0 & 0 & \sqrt{\nu  \eta } & 0 & -\sqrt{(1-\nu ) \eta } & 0 \\
 0 & 0 & 0 & \sqrt{(1-\nu ) (1-\eta )} & 0 & \sqrt{\eta  (1-\nu )+\nu } & 0 & 0 \\
 0 & 0 & 0 & 0 & \sqrt{1-\nu } & 0 & \sqrt{\nu } & 0 \\
 \sqrt{(1-\nu ) (1-\eta )} & 0 & 0 & 0 & 0 & 0 & 0 & \frac{\eta  (1-\nu )+\nu }{\sqrt{1-(1-\nu ) (1-\eta )}} \\
\end{array}
\right)} 
\label{usb_1} \\
U_{SB}^{(2)} &=& \tiny{\left(
\begin{array}{cccccccc}
 \sqrt{\nu } & 0 & -\frac{\sqrt{\eta  (1-\nu )} \sqrt{\nu }}{\sqrt{1-\eta  (1-\nu )}} & 0 & 0 & 0 & 0 & -\frac{\sqrt{(1-\eta ) (1-\nu )}}{\sqrt{1-\eta  (1-\nu )}} \\
 0 & \frac{\sqrt{\eta  \nu }}{\sqrt{1-(1-\nu )}} & 0 & 0 & \sqrt{(1-\eta ) \nu } & 0 & -\sqrt{1-\nu } \sqrt{1-\eta } & 0 \\
 \sqrt{\eta  (1-\nu )} & 0 & \sqrt{1-\eta  (1-\nu )} & 0 & 0 & 0 & 0 & 0 \\
 0 & 0 & 0 & \sqrt{\eta  (1-\nu )+\nu } & 0 & -\sqrt{(1-\eta ) (1-\nu )} & 0 & 0 \\
 0 & -\frac{\sqrt{(1-\eta ) \nu }}{\sqrt{1-(1-\nu )}} & 0 & 0 & \sqrt{\eta  \nu } & 0 & -\sqrt{1-\nu } \sqrt{\eta } & 0 \\
 0 & 0 & 0 & \sqrt{(1-\eta ) (1-\nu )} & 0 & \sqrt{\eta  (1-\nu )+\nu } & 0 & 0 \\
 0 & 0 & 0 & 0 & \sqrt{1-\nu } & 0 & \sqrt{\nu } & 0 \\
 \sqrt{(1-\eta ) (1-\nu )} & 0 & -\frac{(1-\nu ) \sqrt{\eta  (1-\eta )}}{\sqrt{1-\eta  (1-\nu )}} & 0 & 0 & 0 & 0 & \frac{\sqrt{\nu }}{\sqrt{1-\eta  (1-\nu )}} \\
\end{array}
\right)}.
\label{usb_2}
\end{eqnarray}
\end{widetext}
Note that, in these cases the aforesaid GADC appears as a dynamical process in time (as for example, by solving the optical master equation with the initial state of the heat bath taken to be squeezed vacuum). Our aim is to find out the Kraus operator representation of the quantum channel whose unitary dilation corresponds to the inverse of this unitary evolution.
As $U_{SB}$ is unitary, we must have $U_{SB}^{-1}=U_{SB}^\dagger$. Now under the action of the inverse unitary evolution $U_{SB}^{-1}$, the state of the system at the output end is given by, $Tr_{B} [U_{SB}^\dagger (\sigma_{S} \otimes \ket{1}_{B} \bra{1}) U_{SB}]$, where $\sigma_{S}$ is the state of the system just before the action of the inverse unitary. Now if $J_{i}$ for $i=1,2,3,4$ be the Kraus operators corresponding to the channel described by the inverse unitary, one must have, 
\begin{eqnarray}
Tr_{B} [U_{SB}^\dagger (\sigma_{S} \otimes \ket{1}_{B} \bra{1}) U_{SB}]= \sum_{i=1}^{4} J_{i} \sigma_{S} J_{i}^{\dagger},
\label{inv_uni_evolution}
\end{eqnarray}
with, $J_{i}= {}_B \bra{i} U_{SB}^{-1} \ket{1}_{B}$ for $i=1,2,3,4$. In fact, if $U_{SB}=\sum_{k,l=1}^{2} \sum_{\alpha, \beta=1}^{4} u_{k \alpha, l \beta} \ket{k}_{S}\!\bra{l} \otimes \ket{\alpha}_{B}\!\bra{\beta}$, then $U_{SB}^{-1}=U_{SB}^{\dagger}=\sum_{k,l=1}^{2} \sum_{\alpha, \beta=1}^{4} u_{k \alpha, l \beta}^* \ket{l}_{S}\!\bra{k} \otimes \ket{\beta}_{B}\!\bra{\alpha}$, and so,
\begin{eqnarray}
J_i &=& \sum_{k,l=1}^{2} \sum_{\alpha, \beta=1}^{4} u_{k \alpha, l \beta}^* \braket{i|\beta}_{B} \braket{\alpha |1}_{B} \ket{l}_{S}\!\bra{k} \nonumber\\
&=& \sum_{k,l=1}^{2} u_{k 1, l i}^* \ket{l}_{S}\!\bra{k}~~ for~ i=1,2,3,4.
\label{gen_kraus_inv}
\end{eqnarray}
Using the technique mentioned above, we find out the Kraus operators ($J_{1}$, $J_{2}$, $J_{3}$, $J_{4}$) corresponding to the channel which is given by the inverse of the unitary $U_{SB}$.

\subsection{Fidelity of the quantum state}

We can now consider the entire scenario in the following fashion, 
\begin{itemize}
\item[$\rho_{AB}$]  $\xrightarrow{GADC} \rho_{AB}^{\prime} =  \sum_{i=1}^{4} (\mathbb{I} \otimes K_{i}) \rho_{AB} (\mathbb{I} \otimes {K_{i}}^\dagger)$.
\item[$\rho_{AB}$]  $\xrightarrow{GADC} \rho_{AB}^{\prime} =  \sum_{i=1}^{4} (\mathbb{I} \otimes K_{i}) \rho_{AB} (\mathbb{I} \otimes {K_{i}}^\dagger)\xrightarrow{\mathcal{M_N}}  \rho_{AB}^{\prime \prime}$.
\end{itemize}

Here, $\mathcal{M_N}$ represents the action of the inverse map on the evolved state under environmental interaction. Basically, in this case, we consider a two qubit ancilla along with the initial quantum state, and let the whole 4-qubit state pass through the unitary dilation corresponding to the GADC in the following way,
\begin{eqnarray}
\rho_{AB} \xrightarrow{GADC} \rho_{AB}^{\prime} = Tr_{B^{\prime} B^{\prime \prime}} &[& (\mathbb{I}_A \otimes U_{SB}) (\rho_{AB} \otimes \phi_{B^{\prime} B^{\prime \prime}}) \nonumber \\
&&(\mathbb{I}_A \otimes U_{SB})^{\dagger}].
\label{uni_evo}
\end{eqnarray}
After obtaining the evolved state by tracing out the ancillary qubits, the process of initialization of ancilla is employed and let the whole system pass through the inverse unitary map which corresponds to the reverse effect of the channel action. In this case 
one may check the closeness of the state $\rho_{AB}$ with $\rho_{AB}^{\prime \prime}$ and compare it with the closeness of states $\rho_{AB}$ and$\rho_{AB}^{\prime}$. The fidelity as a quantifier for closeness, is given as,
\begin{eqnarray}
F(\rho_i,\rho_f) \equiv Tr[\sqrt{(\rho_i)^{\frac{1}{2}} \rho_f (\rho_i)^{\frac{1}{2}}}].
\label{fidelity}
\end{eqnarray}
For the approach of preservation of entanglement and steerability to be successful,  implementation of the inverse map on the evolved system's state and the initialized ancilla, should lead to a high fidelity for a large  range of values of the damping channel parameter.

\subsection{To preserve quantum correlation using unitary dilation}

\begin{figure*}[ht]
\resizebox{14cm}{8cm}{\includegraphics{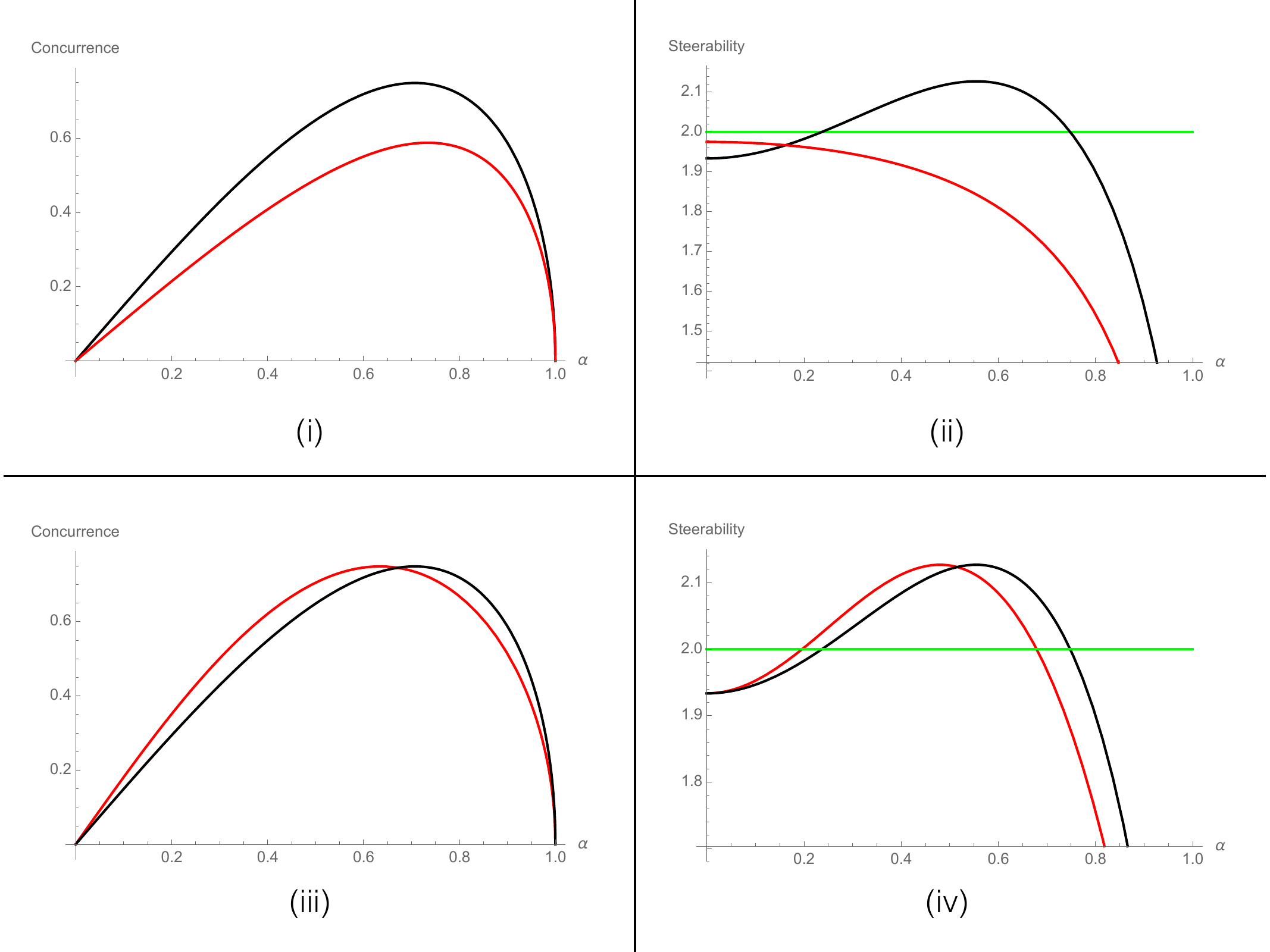}}
\caption{\footnotesize (Colour online) For all the plots, the channel parameters are fixed at, $\nu=0.051$ and $\eta=0.673$ and the initial state is taken as the parallel Bell state i.e. $\ket{\phi^{+}}\bra{\phi^{+}}$. (i) and (ii) are the cases when the weak measurement technique is  applied to manage the effect of GADC. (iii) and (iv) are the cases when the newly proposed technique (POVM corresponding to $U_{SB}^{(1)}$ applied \textbf{before} the channel action) is employed.  In all the cases, black curves denote the quantity under the environmental interaction only and red curves denote when the respective technique being employed. Green line corresponds to the limit of the violation of the ACHSH inequality. Note that y-axes of (ii) and (iv) are of varying scales.}
\label{plot_before_genPOVM1}
\end{figure*}



\begin{figure*}[ht]
\resizebox{14cm}{8cm}{\includegraphics{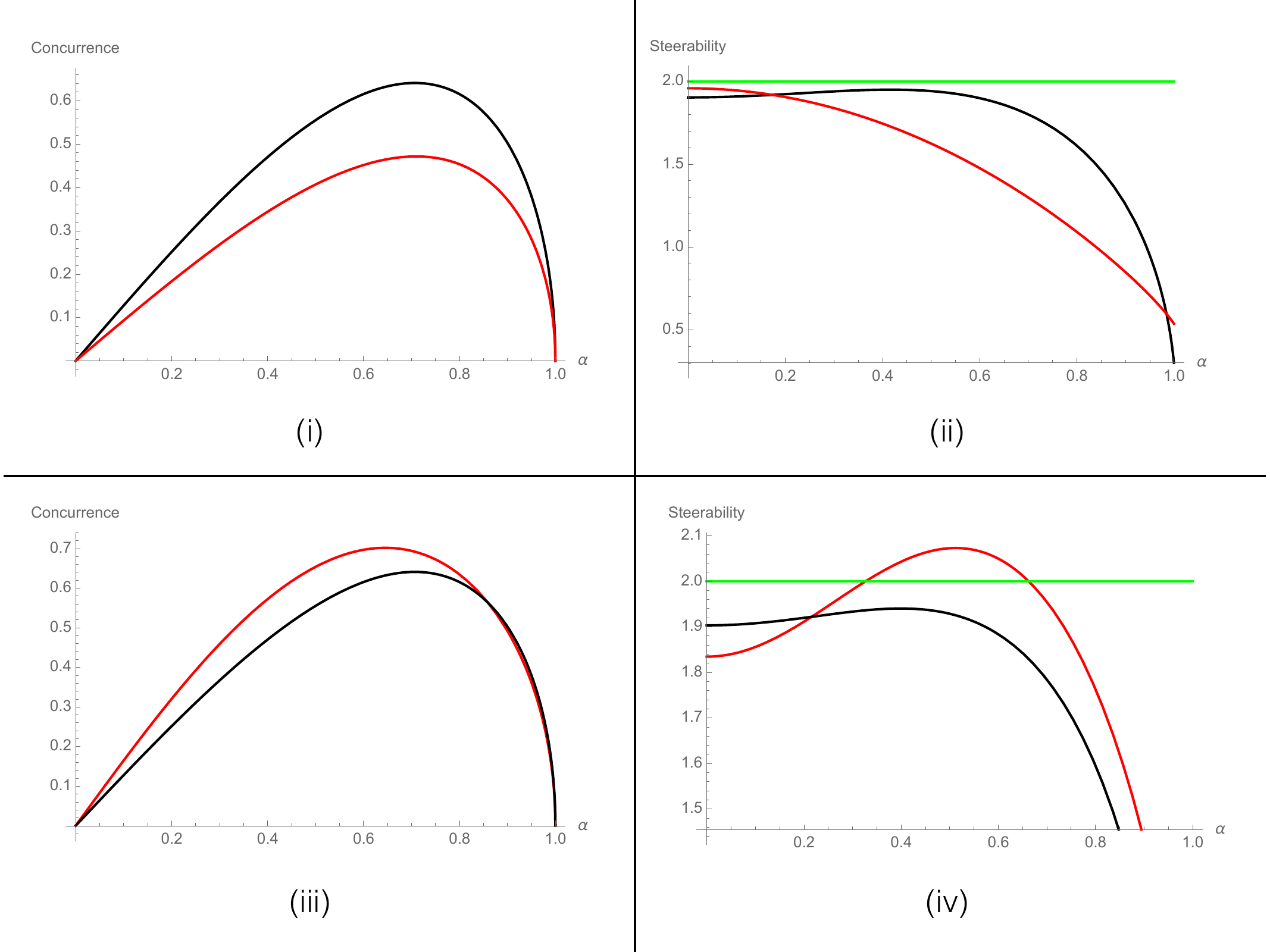}}
\caption{\footnotesize (Colour online) For all the plots, the channel parameters are fixed at, $\nu=0.054$ and $\eta=0.551$ and the initial state is taken as the parallel Bell state i.e. $\ket{\phi^+}\bra{\phi^+}$. (i) and (ii) are the cases when the weak measurement technique is  applied to manage the effect of GADC. (iii) and (iv) are the cases when the newly proposed technique (POVM corresponding to $U_{SB}^{(1)}$ applied \textbf{after} the channel action) is employed. In all the cases, black curves denote the quantity under the environmental interaction only and red curves denote when the respective technique being employed. Green line corresponds to the limit of the violation of the ACHSH inequality. Note that y-axes of (i) and (iii) are of varying scales and similarly for (ii) and (iv).}
\label{plot_after_genPOVM1} 

\resizebox{14cm}{8cm}{\includegraphics{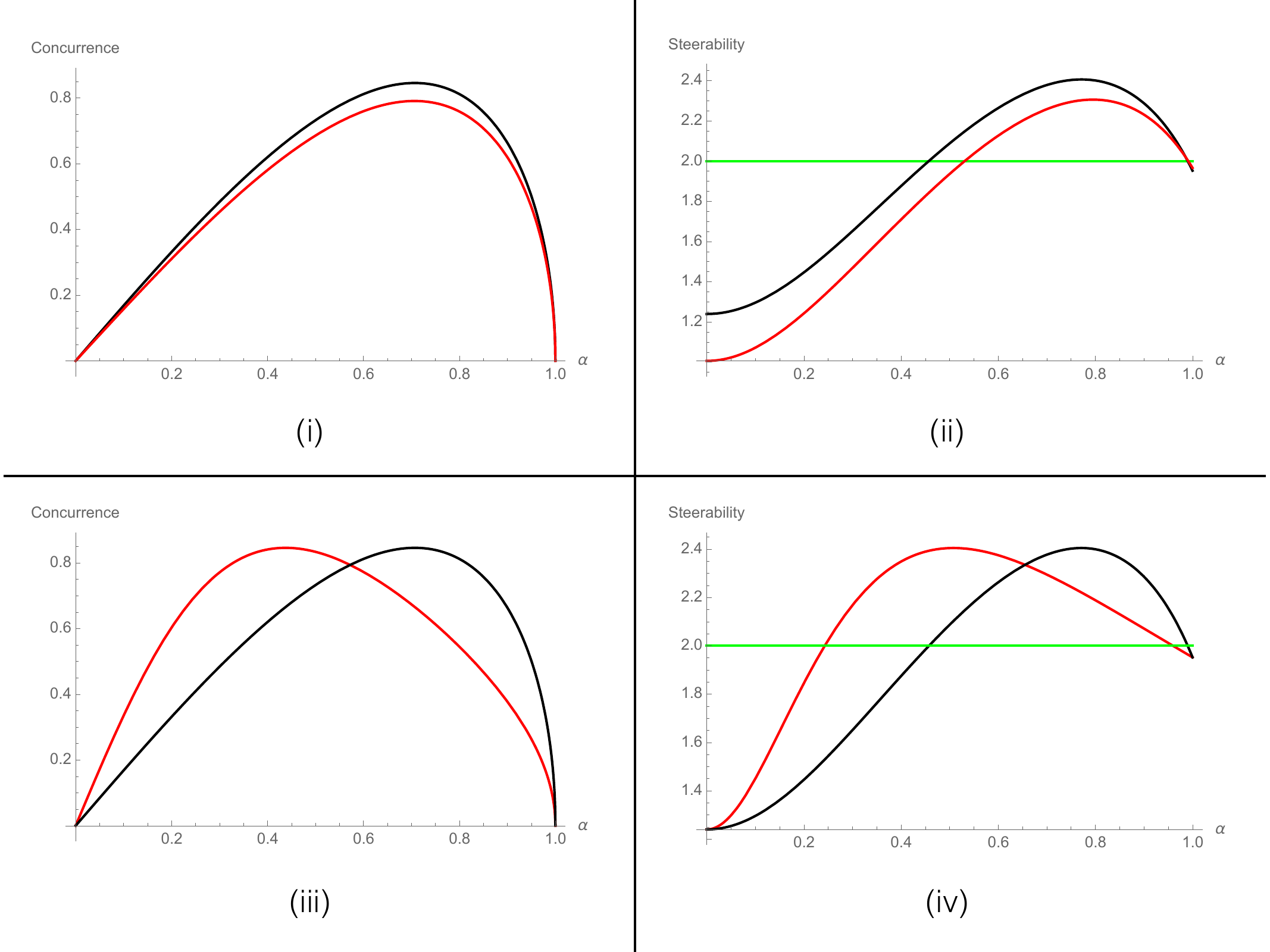}}
\caption{\footnotesize (Colour online) For all the plots, the channel parameters are fixed at, $\nu=0.059$ and $\eta=0.798$ and the initial state is taken as the antiparallel Bell state i.e. $\ket{\psi^+}\bra{\psi^+}$. (i) and (ii) are the cases when the weak measurement technique is  applied to manage the effect of GADC. (iii) and (iv) are the cases when the newly proposed technique (POVM corresponding to $U_{SB}^{(2)}$ applied \textbf{before} the channel action) is employed. In all the cases, black curves denote the quantity under the environmental interaction only and red curves denote when the respective technique being employed. Green line corresponds to the limit of the violation of the ACHSH inequality. Note that y-axes of (ii) and (iv) are of varying scales.}
\label{plot_before_genPOVM2}
\end{figure*}

Following the procedure, mentioned above, of finding the Kraus operator representation associated to the inverse of the unitary dilation of a quantum channel, here we obtain the set of Kraus operators $\lbrace J_{1}^{(1)}, J_{2}^{(1)}, J_{3}^{(1)}, J_{4}^{(1)} \rbrace$ and $\lbrace J_{1}^{(2)}, J_{2}^{(2)}, J_{3}^{(2)}, J_{4}^{(2)} \rbrace$ corresponding to the unitaries given in Eq. (\ref{usb_1}) and (\ref{usb_2}) respectively, which are illustrated below. 
\begin{eqnarray}
&J_{1}^{(1)}&=\left(
\begin{array}{cc}
 \sqrt{\nu } & 0 \\
 0 & \sqrt{\eta  \nu } \\
\end{array}
\right) \nonumber \\
&J_{2}^{(1)}&=\left(
\begin{array}{cc}
 -\frac{\sqrt{\eta -\eta  \nu }}{\sqrt{-\nu  \eta +\eta +\nu }} & 0 \\
 0 & -\sqrt{\eta -\eta  \nu } \\
\end{array}
\right) \nonumber \\
&J_{3}^{(1)}&=\left(
\begin{array}{cc}
 0 & -\sqrt{1-\eta } \\
 0 & 0 \\
\end{array}
\right) \nonumber \\
&J_{4}^{(1)}&=\left(
\begin{array}{cc}
 0 & 0 \\
 -\frac{\sqrt{(\eta -1) (\nu -1) \nu }}{\sqrt{-\nu  \eta +\eta +\nu }} & 0 \\
\end{array}
\right),
\label{Kraus_usb_1_inv}
\end{eqnarray}
and,
\begin{eqnarray}
&J_{1}^{(2)}&=\left(
\begin{array}{cc}
 \sqrt{\nu } & 0 \\
 0 & \sqrt{\eta  \nu } \\
\end{array}
\right) \nonumber \\
&J_{2}^{(2)}&=\left(
\begin{array}{cc}
 -\frac{\sqrt{\nu } \sqrt{\eta -\eta  \nu }}{\sqrt{\eta  (\nu -1)+1}} & 0 \\
 0 & -\sqrt{\eta } \sqrt{1-\nu } \\
\end{array}
\right) \nonumber \\
&J_{3}^{(2)}&=\left(
\begin{array}{cc}
 0 & -\sqrt{1-\eta } \\
 0 & 0 \\
\end{array}
\right) \nonumber \\
&J_{4}^{(2)}&=\left(
\begin{array}{cc}
 0 & 0 \\
 -\frac{\sqrt{(\eta -1) (\nu -1)}}{\sqrt{\eta  (\nu -1)+1}} & 0 \\
\end{array}
\right).
\label{Kraus_usb_2_inv}
\end{eqnarray}
Note that, $\sum_{i=1}^{4} {J_i^{(1)}}^\dagger J_i^{(1)} = \openone$ and $\sum_{i=1}^{4} {J_i^{(2)}}^\dagger J_i^{(2)} = \openone$. Let us now consider that one side ($B$, say) of the bipartite system $AB$ is interacting with the environment through a GADC and hence we apply the selective POVM constructed from the individual element of the Kraus representaion. In this scenario, we consider two different cases depending upon the order of application of the POVM. \\
\textit{Case I}:
\begin{itemize}
\item[$\rho_{AB}$]  $\xrightarrow{GADC} \rho_{AB}^{\prime} =  \sum_{i=1}^{4} (\openone \otimes K_{i}) \rho_{AB} (\openone \otimes {K_{i}}^\dagger)$.
\item[$\rho_{AB}$]  $\xrightarrow{POVM} \rho_{AB}^{p (i)} = \frac{(\openone \otimes {\lbrace J_i^\dagger J_i \rbrace}^{\frac{1}{2}}) \rho_{AB} (\openone \otimes {{\lbrace J_i^\dagger J_i \rbrace}^{\frac{1}{2}}}^\dagger)}{Tr[(\openone \otimes {\lbrace J_i^\dagger J_i \rbrace}^{\frac{1}{2}}) \rho_{AB} (\openone \otimes {{\lbrace J_i^\dagger J_i \rbrace}^{\frac{1}{2}}}^\dagger)]}$ \\
$ \xrightarrow{GADC} \rho_{AB}^{pd (i)} = \sum_{i=1}^{4} (\openone \otimes K_{i}) \rho_{AB}^{p} (\openone \otimes {K_{i}}^\dagger)$. 
\end{itemize}
\textit{Case II}:
\begin{itemize}
\item[$\rho_{AB}$]  $\xrightarrow{GADC} \rho_{AB}^{\prime} =  \sum_{i=1}^{4} (\openone \otimes K_{i}) \rho_{AB} (\openone \otimes {K_{i}}^\dagger)$.
\item[$\rho_{AB}$]  $\xrightarrow{GADC} \rho_{AB}^{\prime} = \sum_{i=1}^{4} (\openone \otimes K_{i}) \rho_{AB}^{p} (\openone \otimes {K_{i}}^\dagger)$ \\
$\xrightarrow{POVM} \rho_{AB}^{dp (i)} = \frac{(\openone \otimes {\lbrace J_i^\dagger J_i \rbrace}^{\frac{1}{2}}) \rho_{AB}^{\prime} (\openone \otimes {{\lbrace J_i^\dagger J_i \rbrace}^{\frac{1}{2}}}^\dagger)}{Tr[(\openone \otimes {\lbrace J_i^\dagger J_i \rbrace}^{\frac{1}{2}}) \rho_{AB}^{\prime} (\openone \otimes {{\lbrace J_i^\dagger J_i \rbrace}^{\frac{1}{2}}}^\dagger)]}$. 
\end{itemize}
For Case I, we comparatively study the concurrence and steerability of the states $\rho_{AB}^{\prime}$ and $\rho_{AB}^{pd (i)}$, whereas in Case II, the similar protocol has been followed for the states $\rho_{AB}^{\prime}$ and $\rho_{AB}^{dp (i)}$. 
In the figures (\ref{plot_before_genPOVM1}), (\ref{plot_after_genPOVM1}) and (\ref{plot_before_genPOVM2}), we provide plots of the concurrence and steerabilty 
illustrating examples of the above two cases taking initial Bell states. Comparisons with
the weak measurement technique discuused earlier, is also provided for the corresponding
values of channel parameters.

\begin{figure*}[ht]
\resizebox{14cm}{8cm}{\includegraphics{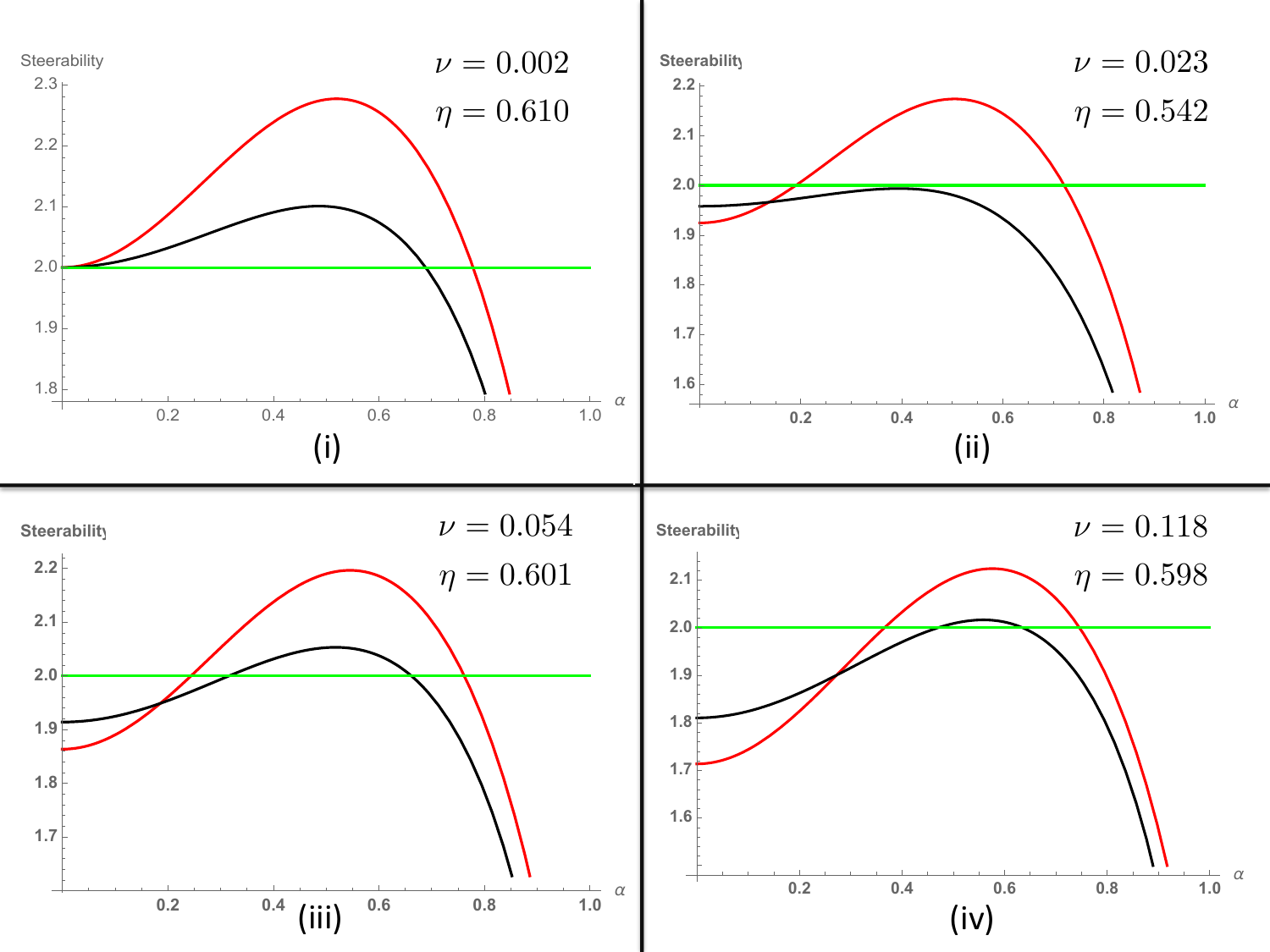}}
\caption{\footnotesize (Colour online) The comparison of steerability in terms of the violation ACHSH inequality for different values of $\nu$ and $\eta$ (the damping parameters of the given channel), when the initial state is considered to be a Bell parallel state, $\ket{\phi^{+}}\bra{\phi^{+}}$. POVM corresponding to $U_{SB}^{(1)}$ is applied \textbf{after} the channel action. The plot colors have their usual meaning.}
\label{plot_after_genPOVM1_nueta}
\end{figure*}

From the plots, improvement of quantum correlations on the application of POVM can be seen, over the sole interaction with the environment through GADC. This is a more general approach than the approach discussed in the Sec (\ref{sec_weak}) and improvement can be seen over a larger range of values of the damping parameters $\eta$ and $\nu$ in this case, compared to the weak measurement technique. Also, finding the suitable unitary matrix $U_{SB}$ just by the method of basis expansion, one can identify the helpful POVM in protecting the quantum correlation, for a particular damping channel. From Figure (\ref{plot_after_genPOVM1_nueta}), the improvement in steerability can be seen for several different values of the channel parameters. It can be observed that the unitary  $U_{SB}^{(1)}$  applied after the environmental interaction works well for the chosen set of values of $\nu$ and $\eta$ and for initial state being the parallel Bell state, i.e. $\ket{\phi^{+}}\bra{\phi^{+}}$.
   On the other hand, $U_{SB}^{(2)}$ can be chosen for an effective solution for the antiparallel Bell state $\ket{\psi^+}\bra{\psi^+}$ when applied before the environmental interaction, as can be seen from Fig (\ref{plot_before_genPOVM2}).

The motivation behind introducing the unitary dilation $U_{SB}$ of a quantum channel $\Lambda$ (acting on $S$) and the quantum channel $\Lambda^{\prime}$ (acting on $S$) whose unitary dilation being $U_{SB}^{-1}$, is to nullify the action of $\Lambda$. Such a scheme will be completely successful provided it can be guaranteed that $\Lambda(\rho_{S}) \otimes \sigma_{B}^{(1)}$ is close to $U_{SB}[\Lambda(\rho_{S}) \otimes \sigma_{B}^{(0)}] U_{SB}^{-1}$ for two fixed states $\sigma_B^{(0)}$ and $\sigma_{B}^{(1)}$ of B. Needless to say,  such a condition can not be satisfied, in general. Hence, our method can only recover the quantum correlation of the state partially. There are decoherence controlling models in literature for the noisy channels obtained by solving the optical master equation for thermal bath \cite{MS_77, VL_98}. But for the particular case of GADC, which can be obtained from a squeezed thermal bath, our method of protecting quantum correlations is an effective pr
 ocedure. 

A possible way to choose the unitary dilation $U_{SB}$ -- satisfying (i) of Sec(\ref{Mot_app}) and approximately satisfying (ii) of Sec (\ref{Mot_app})-- is the following. Choose $U_{SB}$ in such a way that (a) it satisfies (i), (b) the reduced states $Tr_S[U_{SB}({\rho}_S \otimes {\sigma}^{(0)}_B)U_{SB}^{\dagger}]$ of $B$ are closed to each other over all choices of the initial states $\rho_S$ of S, and (c) $U_{SB}$ has minimal entangling power \cite{ZZF_00}. Note that conditions (b) and (c) together (which are not necessarily independent of each other) satisfy condition (ii) above. Also note that, the aforesaid ancilla state ${\sigma}^1_B$ may now be chosen (given that $U_{SB}$ satisfies conditions (a), (b), and (c)) to be the average of $Tr_S[U_{SB}({\rho}_S \otimes {\sigma}^{(0)}_B)U_{SB}^{\dagger}]$  over all possible choices of the initial states ${\sigma}_S$ of the system $S$. Needless to say that finding out the unitary dilation $U_{SB}$, satisfying the conditions (a)
  - (c), is computationally a challenging task although, as a method, this has a universal character.

\section{Conclusions}
\label{concl}

In this paper, we have studied  the problem of preserving quantum correlations that are useful in different information processing tasks, under the action of a noisy environment. Here, we choose GADC as the environmental noise and check its effect on entanglement and steerability of an initial pure bipartite state. First, we have employed the technique of weak measurement and reversal and found that a certain amount of improvement could be achieved. But, it can also be seen that this improvement is restricted to some particular values of the damping parameters of the corresponding channel. 

We have next introduced another method for the preservation of correlations using a unitary dilation of the operator sum representation of the channel. Interestingly, as the choice of the unitary is not unique, it provides us the freedom to choose the inverse evolution of the unitary, and hence the Kraus operators according to our convenience. Choosing different unitaries and consequently their inverses gives us the scope to extend our scheme over a larger range of the damping parameters, thus improving the quality of preservation. Note that in the present paper we have dealt with two particular unitaries corresponding the Kraus representation of GADC for the illustration of our approach. However, it is possible to construct other unitaries taking the conditions of orthogonality and normalisation into account. Also, there can be other protocols using  partial inversion of CPTP channels leading to improvement in terms of fidelity \cite{KBF_20}.

As a future direction, this method can be applied to other noisy channels and the choice of the unitary can be made suitably in order to generate an optimal scheme for protecting quantum correlations under the action of different noisy environments. On the other hand, another interesting and useful approach towards the preservation of correlations would be to find the optimized solution for the problem. It may be noted that the considered prescription gives us a wide opportunity to choose between numerous unitary dilation corresponding to a given channel as it is not unique. Hence, it could be more helpful to choose the unitary corresponding to which the  highest fidelity can be obtained for a particular scenario. Also, one has to keep in mind that the ancillary state that would be traced out after the unitary evolution, should be as near as possible to the initially taken ancillary state. In this way we would introduce the least amount of correlation to the ancillary system 
 which gives us the opportunity to find the unitary that performs better in preserving the correlation in the main system under consideration.

\begin{acknowledgments}
SG1 would like to acknowledge Sagnik Chakraborty and Samyadeb Bhattacharya for helpful discussions. SG2 would like to thankfully acknowledge the hospitality of S. N. Bose National Centre for Basic Sciences during his visits in the recent past during which part of the work was done. ASM acknowledges Project no. DST/ICPS/QuEST/2018/98 of Department of Science and
Technology, Government of India.
\end{acknowledgments}

\bibliography{GADC.bib}

\begin{thebibliography}{63}%
\makeatletter
\providecommand \@ifxundefined [1]{%
 \@ifx{#1\undefined}
}%
\providecommand \@ifnum [1]{%
 \ifnum #1\expandafter \@firstoftwo
 \else \expandafter \@secondoftwo
 \fi
}%
\providecommand \@ifx [1]{%
 \ifx #1\expandafter \@firstoftwo
 \else \expandafter \@secondoftwo
 \fi
}%
\providecommand \natexlab [1]{#1}%
\providecommand \enquote  [1]{``#1''}%
\providecommand \bibnamefont  [1]{#1}%
\providecommand \bibfnamefont [1]{#1}%
\providecommand \citenamefont [1]{#1}%
\providecommand \href@noop [0]{\@secondoftwo}%
\providecommand \href [0]{\begingroup \@sanitize@url \@href}%
\providecommand \@href[1]{\@@startlink{#1}\@@href}%
\providecommand \@@href[1]{\endgroup#1\@@endlink}%
\providecommand \@sanitize@url [0]{\catcode `\\12\catcode `\$12\catcode
  `\&12\catcode `\#12\catcode `\^12\catcode `\_12\catcode `\%12\relax}%
\providecommand \@@startlink[1]{}%
\providecommand \@@endlink[0]{}%
\providecommand \url  [0]{\begingroup\@sanitize@url \@url }%
\providecommand \@url [1]{\endgroup\@href {#1}{\urlprefix }}%
\providecommand \urlprefix  [0]{URL }%
\providecommand \Eprint [0]{\href }%
\providecommand \doibase [0]{http://dx.doi.org/}%
\providecommand \selectlanguage [0]{\@gobble}%
\providecommand \bibinfo  [0]{\@secondoftwo}%
\providecommand \bibfield  [0]{\@secondoftwo}%
\providecommand \translation [1]{[#1]}%
\providecommand \BibitemOpen [0]{}%
\providecommand \bibitemStop [0]{}%
\providecommand \bibitemNoStop [0]{.\EOS\space}%
\providecommand \EOS [0]{\spacefactor3000\relax}%
\providecommand \BibitemShut  [1]{\csname bibitem#1\endcsname}%
\let\auto@bib@innerbib\@empty
\bibitem [{\citenamefont {Bennett}\ \emph {et~al.}(1993)\citenamefont
  {Bennett}, \citenamefont {Brassard}, \citenamefont {Cr{\'e}peau},
  \citenamefont {Jozsa}, \citenamefont {Peres},\ and\ \citenamefont
  {Wootters}}]{BBCJPW_93}%
  \BibitemOpen
  \bibfield  {author} {\bibinfo {author} {\bibfnamefont {C.~H.}\ \bibnamefont
  {Bennett}}, \bibinfo {author} {\bibfnamefont {G.}~\bibnamefont {Brassard}},
  \bibinfo {author} {\bibfnamefont {C.}~\bibnamefont {Cr{\'e}peau}}, \bibinfo
  {author} {\bibfnamefont {R.}~\bibnamefont {Jozsa}}, \bibinfo {author}
  {\bibfnamefont {A.}~\bibnamefont {Peres}}, \ and\ \bibinfo {author}
  {\bibfnamefont {W.~K.}\ \bibnamefont {Wootters}},\ }\href@noop {} {\bibfield
  {journal} {\bibinfo  {journal} {Physical review letters}\ }\textbf {\bibinfo
  {volume} {70}},\ \bibinfo {pages} {1895} (\bibinfo {year}
  {1993})}\BibitemShut {NoStop}%
\bibitem [{\citenamefont {Bennett}\ and\ \citenamefont
  {Wiesner}(1992)}]{BW_92}%
  \BibitemOpen
  \bibfield  {author} {\bibinfo {author} {\bibfnamefont {C.~H.}\ \bibnamefont
  {Bennett}}\ and\ \bibinfo {author} {\bibfnamefont {S.~J.}\ \bibnamefont
  {Wiesner}},\ }\href@noop {} {\bibfield  {journal} {\bibinfo  {journal}
  {Physical review letters}\ }\textbf {\bibinfo {volume} {69}},\ \bibinfo
  {pages} {2881} (\bibinfo {year} {1992})}\BibitemShut {NoStop}%
\bibitem [{\citenamefont {Shor}(1995)}]{S_95}%
  \BibitemOpen
  \bibfield  {author} {\bibinfo {author} {\bibfnamefont {P.~W.}\ \bibnamefont
  {Shor}},\ }\href@noop {} {\bibfield  {journal} {\bibinfo  {journal} {Phys.
  Rev. A}\ }\textbf {\bibinfo {volume} {52}},\ \bibinfo {pages} {R2493}
  (\bibinfo {year} {1995})}\BibitemShut {NoStop}%
\bibitem [{\citenamefont {Bennett}\ and\ \citenamefont
  {Brassard}(1984)}]{BB_84}%
  \BibitemOpen
  \bibfield  {author} {\bibinfo {author} {\bibfnamefont {C.~H.}\ \bibnamefont
  {Bennett}}\ and\ \bibinfo {author} {\bibfnamefont {G.}~\bibnamefont
  {Brassard}}\ }(\bibinfo {year} {1984})\ pp.\ \bibinfo {pages}
  {175--179}\BibitemShut {NoStop}%
\bibitem [{\citenamefont {Ekert}(1991)}]{E_91}%
  \BibitemOpen
  \bibfield  {author} {\bibinfo {author} {\bibfnamefont {A.~K.}\ \bibnamefont
  {Ekert}},\ }\href@noop {} {\bibfield  {journal} {\bibinfo  {journal} {Phys.
  Rev. Lett.}\ }\textbf {\bibinfo {volume} {67}},\ \bibinfo {pages} {661}
  (\bibinfo {year} {1991})}\BibitemShut {NoStop}%
\bibitem [{\citenamefont {Branciard}\ \emph {et~al.}(2012)\citenamefont
  {Branciard}, \citenamefont {Cavalcanti}, \citenamefont {Walborn},
  \citenamefont {Scarani},\ and\ \citenamefont {Wiseman}}]{BCWSW_12}%
  \BibitemOpen
  \bibfield  {author} {\bibinfo {author} {\bibfnamefont {C.}~\bibnamefont
  {Branciard}}, \bibinfo {author} {\bibfnamefont {E.~G.}\ \bibnamefont
  {Cavalcanti}}, \bibinfo {author} {\bibfnamefont {S.~P.}\ \bibnamefont
  {Walborn}}, \bibinfo {author} {\bibfnamefont {V.}~\bibnamefont {Scarani}}, \
  and\ \bibinfo {author} {\bibfnamefont {H.~M.}\ \bibnamefont {Wiseman}},\
  }\href {\doibase 10.1103/PhysRevA.85.010301} {\bibfield  {journal} {\bibinfo
  {journal} {Phys. Rev. A}\ }\textbf {\bibinfo {volume} {85}},\ \bibinfo
  {pages} {010301 (R)} (\bibinfo {year} {2012})}\BibitemShut {NoStop}%
\bibitem [{\citenamefont {Bru{\ss}}\ \emph {et~al.}(2000)\citenamefont
  {Bru{\ss}}, \citenamefont {Faoro}, \citenamefont {Macchiavello},\ and\
  \citenamefont {Palma}}]{BFMP_00}%
  \BibitemOpen
  \bibfield  {author} {\bibinfo {author} {\bibfnamefont {D.}~\bibnamefont
  {Bru{\ss}}}, \bibinfo {author} {\bibfnamefont {L.}~\bibnamefont {Faoro}},
  \bibinfo {author} {\bibfnamefont {C.}~\bibnamefont {Macchiavello}}, \ and\
  \bibinfo {author} {\bibfnamefont {G.~M.}\ \bibnamefont {Palma}},\ }\href@noop
  {} {\bibfield  {journal} {\bibinfo  {journal} {Journal of Modern Optics}\
  }\textbf {\bibinfo {volume} {47}},\ \bibinfo {pages} {325} (\bibinfo {year}
  {2000})}\BibitemShut {NoStop}%
\bibitem [{\citenamefont {Schumacher}\ and\ \citenamefont
  {Nielsen}(1996)}]{SN_96}%
  \BibitemOpen
  \bibfield  {author} {\bibinfo {author} {\bibfnamefont {B.}~\bibnamefont
  {Schumacher}}\ and\ \bibinfo {author} {\bibfnamefont {M.~A.}\ \bibnamefont
  {Nielsen}},\ }\href {\doibase 10.1103/PhysRevA.54.2629} {\bibfield  {journal}
  {\bibinfo  {journal} {Phys. Rev. A}\ }\textbf {\bibinfo {volume} {54}},\
  \bibinfo {pages} {2629} (\bibinfo {year} {1996})}\BibitemShut {NoStop}%
\bibitem [{\citenamefont {Srikanth}\ and\ \citenamefont
  {Banerjee}(2008)}]{SB_08}%
  \BibitemOpen
  \bibfield  {author} {\bibinfo {author} {\bibfnamefont {R.}~\bibnamefont
  {Srikanth}}\ and\ \bibinfo {author} {\bibfnamefont {S.}~\bibnamefont
  {Banerjee}},\ }\href@noop {} {\bibfield  {journal} {\bibinfo  {journal}
  {Phys. Rev. A}\ }\textbf {\bibinfo {volume} {77}},\ \bibinfo {pages} {012318}
  (\bibinfo {year} {2008})}\BibitemShut {NoStop}%
\bibitem [{\citenamefont {Fujiwara}(2004)}]{F_04}%
  \BibitemOpen
  \bibfield  {author} {\bibinfo {author} {\bibfnamefont {A.}~\bibnamefont
  {Fujiwara}},\ }\href@noop {} {\bibfield  {journal} {\bibinfo  {journal}
  {Phys. Rev. A}\ }\textbf {\bibinfo {volume} {70}},\ \bibinfo {pages} {012317}
  (\bibinfo {year} {2004})}\BibitemShut {NoStop}%
\bibitem [{\citenamefont {Nielsen}\ and\ \citenamefont {Chuang}(2002)}]{NC_02}%
  \BibitemOpen
  \bibfield  {author} {\bibinfo {author} {\bibfnamefont {M.~A.}\ \bibnamefont
  {Nielsen}}\ and\ \bibinfo {author} {\bibfnamefont {I.}~\bibnamefont
  {Chuang}},\ }\href@noop {} {\enquote {\bibinfo {title} {Quantum computation
  and quantum information},}\ } (\bibinfo {year} {2002})\BibitemShut {NoStop}%
\bibitem [{\citenamefont {Streltsov}\ \emph {et~al.}(2011)\citenamefont
  {Streltsov}, \citenamefont {Kampermann},\ and\ \citenamefont
  {Bru{\ss}}}]{SKB_11}%
  \BibitemOpen
  \bibfield  {author} {\bibinfo {author} {\bibfnamefont {A.}~\bibnamefont
  {Streltsov}}, \bibinfo {author} {\bibfnamefont {H.}~\bibnamefont
  {Kampermann}}, \ and\ \bibinfo {author} {\bibfnamefont {D.}~\bibnamefont
  {Bru{\ss}}},\ }\href@noop {} {\bibfield  {journal} {\bibinfo  {journal}
  {Phys. Rev. Lett.}\ }\textbf {\bibinfo {volume} {107}},\ \bibinfo {pages}
  {170502} (\bibinfo {year} {2011})}\BibitemShut {NoStop}%
\bibitem [{\citenamefont {Badziag}\ \emph {et~al.}(2000)\citenamefont
  {Badziag}, \citenamefont {Horodecki}, \citenamefont {Horodecki},\ and\
  \citenamefont {Horodecki}}]{BHHH_00}%
  \BibitemOpen
  \bibfield  {author} {\bibinfo {author} {\bibfnamefont {P.}~\bibnamefont
  {Badziag}}, \bibinfo {author} {\bibfnamefont {M.}~\bibnamefont {Horodecki}},
  \bibinfo {author} {\bibfnamefont {P.}~\bibnamefont {Horodecki}}, \ and\
  \bibinfo {author} {\bibfnamefont {R.}~\bibnamefont {Horodecki}},\ }\href@noop
  {} {\bibfield  {journal} {\bibinfo  {journal} {Phys. Rev. A}\ }\textbf
  {\bibinfo {volume} {62}},\ \bibinfo {pages} {012311} (\bibinfo {year}
  {2000})}\BibitemShut {NoStop}%
\bibitem [{\citenamefont {Bandyopadhyay}(2002)}]{B_02}%
  \BibitemOpen
  \bibfield  {author} {\bibinfo {author} {\bibfnamefont {S.}~\bibnamefont
  {Bandyopadhyay}},\ }\href@noop {} {\bibfield  {journal} {\bibinfo  {journal}
  {Phys. Rev. A}\ }\textbf {\bibinfo {volume} {65}},\ \bibinfo {pages} {022302}
  (\bibinfo {year} {2002})}\BibitemShut {NoStop}%
\bibitem [{\citenamefont {Koashi}\ and\ \citenamefont {Ueda}(1999)}]{KU_99}%
  \BibitemOpen
  \bibfield  {author} {\bibinfo {author} {\bibfnamefont {M.}~\bibnamefont
  {Koashi}}\ and\ \bibinfo {author} {\bibfnamefont {M.}~\bibnamefont {Ueda}},\
  }\href {\doibase 10.1103/PhysRevLett.82.2598} {\bibfield  {journal} {\bibinfo
   {journal} {Phys. Rev. Lett.}\ }\textbf {\bibinfo {volume} {82}},\ \bibinfo
  {pages} {2598} (\bibinfo {year} {1999})}\BibitemShut {NoStop}%
\bibitem [{\citenamefont {Korotkov}\ and\ \citenamefont
  {Jordan}(2006)}]{KJ_06}%
  \BibitemOpen
  \bibfield  {author} {\bibinfo {author} {\bibfnamefont {A.~N.}\ \bibnamefont
  {Korotkov}}\ and\ \bibinfo {author} {\bibfnamefont {A.~N.}\ \bibnamefont
  {Jordan}},\ }\href {\doibase 10.1103/PhysRevLett.97.166805} {\bibfield
  {journal} {\bibinfo  {journal} {Phys. Rev. Lett.}\ }\textbf {\bibinfo
  {volume} {97}},\ \bibinfo {pages} {166805} (\bibinfo {year}
  {2006})}\BibitemShut {NoStop}%
\bibitem [{\citenamefont {Kim}\ \emph {et~al.}(2009)\citenamefont {Kim},
  \citenamefont {Cho}, \citenamefont {Ra},\ and\ \citenamefont {Kim}}]{KC_09}%
  \BibitemOpen
  \bibfield  {author} {\bibinfo {author} {\bibfnamefont {Y.-S.}\ \bibnamefont
  {Kim}}, \bibinfo {author} {\bibfnamefont {Y.-W.}\ \bibnamefont {Cho}},
  \bibinfo {author} {\bibfnamefont {Y.-S.}\ \bibnamefont {Ra}}, \ and\ \bibinfo
  {author} {\bibfnamefont {Y.-H.}\ \bibnamefont {Kim}},\ }\href@noop {}
  {\bibfield  {journal} {\bibinfo  {journal} {Optics express}\ }\textbf
  {\bibinfo {volume} {17}},\ \bibinfo {pages} {11978} (\bibinfo {year}
  {2009})}\BibitemShut {NoStop}%
\bibitem [{\citenamefont {Kim}\ \emph {et~al.}(2012)\citenamefont {Kim},
  \citenamefont {Lee}, \citenamefont {Kwon},\ and\ \citenamefont
  {Kim}}]{KL_12}%
  \BibitemOpen
  \bibfield  {author} {\bibinfo {author} {\bibfnamefont {Y.-S.}\ \bibnamefont
  {Kim}}, \bibinfo {author} {\bibfnamefont {J.-C.}\ \bibnamefont {Lee}},
  \bibinfo {author} {\bibfnamefont {O.}~\bibnamefont {Kwon}}, \ and\ \bibinfo
  {author} {\bibfnamefont {Y.-H.}\ \bibnamefont {Kim}},\ }\href@noop {}
  {\bibfield  {journal} {\bibinfo  {journal} {Nature Physics}\ }\textbf
  {\bibinfo {volume} {8}},\ \bibinfo {pages} {117} (\bibinfo {year}
  {2012})}\BibitemShut {NoStop}%
\bibitem [{\citenamefont {Heibati}\ \emph {et~al.}(2020)\citenamefont
  {Heibati}, \citenamefont {Mani},\ and\ \citenamefont {Karimipour}}]{HMK_20}%
  \BibitemOpen
  \bibfield  {author} {\bibinfo {author} {\bibfnamefont {O.}~\bibnamefont
  {Heibati}}, \bibinfo {author} {\bibfnamefont {A.}~\bibnamefont {Mani}}, \
  and\ \bibinfo {author} {\bibfnamefont {V.}~\bibnamefont {Karimipour}},\
  }\href@noop {} {\bibfield  {journal} {\bibinfo  {journal} {arXiv preprint
  arXiv:2003.01341}\ } (\bibinfo {year} {2020})}\BibitemShut {NoStop}%
\bibitem [{\citenamefont {Pramanik}\ and\ \citenamefont
  {Majumdar}(2013)}]{PM_13}%
  \BibitemOpen
  \bibfield  {author} {\bibinfo {author} {\bibfnamefont {T.}~\bibnamefont
  {Pramanik}}\ and\ \bibinfo {author} {\bibfnamefont {A.}~\bibnamefont
  {Majumdar}},\ }\href@noop {} {\bibfield  {journal} {\bibinfo  {journal}
  {Phys. Lett. A}\ }\textbf {\bibinfo {volume} {377}},\ \bibinfo {pages} {3209}
  (\bibinfo {year} {2013})}\BibitemShut {NoStop}%
\bibitem [{\citenamefont {Misra}\ and\ \citenamefont
  {Sudarshan}(1977)}]{MS_77}%
  \BibitemOpen
  \bibfield  {author} {\bibinfo {author} {\bibfnamefont {B.}~\bibnamefont
  {Misra}}\ and\ \bibinfo {author} {\bibfnamefont {E.~G.}\ \bibnamefont
  {Sudarshan}},\ }\href@noop {} {\bibfield  {journal} {\bibinfo  {journal}
  {Journal of Mathematical Physics}\ }\textbf {\bibinfo {volume} {18}},\
  \bibinfo {pages} {756} (\bibinfo {year} {1977})}\BibitemShut {NoStop}%
\bibitem [{\citenamefont {Facchi}\ \emph {et~al.}(2004)\citenamefont {Facchi},
  \citenamefont {Lidar},\ and\ \citenamefont {Pascazio}}]{FLP_04}%
  \BibitemOpen
  \bibfield  {author} {\bibinfo {author} {\bibfnamefont {P.}~\bibnamefont
  {Facchi}}, \bibinfo {author} {\bibfnamefont {D.}~\bibnamefont {Lidar}}, \
  and\ \bibinfo {author} {\bibfnamefont {S.}~\bibnamefont {Pascazio}},\
  }\href@noop {} {\bibfield  {journal} {\bibinfo  {journal} {Phys. Rev. A}\
  }\textbf {\bibinfo {volume} {69}},\ \bibinfo {pages} {032314} (\bibinfo
  {year} {2004})}\BibitemShut {NoStop}%
\bibitem [{\citenamefont {Viola}\ and\ \citenamefont {Lloyd}(1998)}]{VL_98}%
  \BibitemOpen
  \bibfield  {author} {\bibinfo {author} {\bibfnamefont {L.}~\bibnamefont
  {Viola}}\ and\ \bibinfo {author} {\bibfnamefont {S.}~\bibnamefont {Lloyd}},\
  }\href@noop {} {\bibfield  {journal} {\bibinfo  {journal} {Phys. Rev. A}\
  }\textbf {\bibinfo {volume} {58}},\ \bibinfo {pages} {2733} (\bibinfo {year}
  {1998})}\BibitemShut {NoStop}%
\bibitem [{\citenamefont {Zanardi}(1999)}]{Z_99}%
  \BibitemOpen
  \bibfield  {author} {\bibinfo {author} {\bibfnamefont {P.}~\bibnamefont
  {Zanardi}},\ }\href@noop {} {\bibfield  {journal} {\bibinfo  {journal} {Phys.
  Lett. A}\ }\textbf {\bibinfo {volume} {258}},\ \bibinfo {pages} {77}
  (\bibinfo {year} {1999})}\BibitemShut {NoStop}%
\bibitem [{\citenamefont {Byrd}\ and\ \citenamefont {Lidar}(2002)}]{BL_02}%
  \BibitemOpen
  \bibfield  {author} {\bibinfo {author} {\bibfnamefont {M.~S.}\ \bibnamefont
  {Byrd}}\ and\ \bibinfo {author} {\bibfnamefont {D.~A.}\ \bibnamefont
  {Lidar}},\ }\href@noop {} {\bibfield  {journal} {\bibinfo  {journal} {Quantum
  Information Processing}\ }\textbf {\bibinfo {volume} {1}},\ \bibinfo {pages}
  {19} (\bibinfo {year} {2002})}\BibitemShut {NoStop}%
\bibitem [{\citenamefont {Byrd}\ and\ \citenamefont {Lidar}(2003)}]{BL_03}%
  \BibitemOpen
  \bibfield  {author} {\bibinfo {author} {\bibfnamefont {M.~S.}\ \bibnamefont
  {Byrd}}\ and\ \bibinfo {author} {\bibfnamefont {D.~A.}\ \bibnamefont
  {Lidar}},\ }\href@noop {} {\bibfield  {journal} {\bibinfo  {journal} {Phys.
  Rev. A}\ }\textbf {\bibinfo {volume} {67}},\ \bibinfo {pages} {012324}
  (\bibinfo {year} {2003})}\BibitemShut {NoStop}%
\bibitem [{\citenamefont {Peres}(1980)}]{P_80}%
  \BibitemOpen
  \bibfield  {author} {\bibinfo {author} {\bibfnamefont {A.}~\bibnamefont
  {Peres}},\ }\href@noop {} {\bibfield  {journal} {\bibinfo  {journal}
  {American Journal of Physics}\ }\textbf {\bibinfo {volume} {48}},\ \bibinfo
  {pages} {931} (\bibinfo {year} {1980})}\BibitemShut {NoStop}%
\bibitem [{\citenamefont {Schulman}(1998)}]{S_98}%
  \BibitemOpen
  \bibfield  {author} {\bibinfo {author} {\bibfnamefont {L.}~\bibnamefont
  {Schulman}},\ }\href@noop {} {\bibfield  {journal} {\bibinfo  {journal}
  {Phys. Rev. A}\ }\textbf {\bibinfo {volume} {57}},\ \bibinfo {pages} {1509}
  (\bibinfo {year} {1998})}\BibitemShut {NoStop}%
\bibitem [{\citenamefont {Facchi}\ \emph {et~al.}(2005)\citenamefont {Facchi},
  \citenamefont {Tasaki}, \citenamefont {Pascazio}, \citenamefont {Nakazato},
  \citenamefont {Tokuse},\ and\ \citenamefont {Lidar}}]{FTPNTL_05}%
  \BibitemOpen
  \bibfield  {author} {\bibinfo {author} {\bibfnamefont {P.}~\bibnamefont
  {Facchi}}, \bibinfo {author} {\bibfnamefont {S.}~\bibnamefont {Tasaki}},
  \bibinfo {author} {\bibfnamefont {S.}~\bibnamefont {Pascazio}}, \bibinfo
  {author} {\bibfnamefont {H.}~\bibnamefont {Nakazato}}, \bibinfo {author}
  {\bibfnamefont {A.}~\bibnamefont {Tokuse}}, \ and\ \bibinfo {author}
  {\bibfnamefont {D.}~\bibnamefont {Lidar}},\ }\href@noop {} {\bibfield
  {journal} {\bibinfo  {journal} {Phys. Rev. A}\ }\textbf {\bibinfo {volume}
  {71}},\ \bibinfo {pages} {022302} (\bibinfo {year} {2005})}\BibitemShut
  {NoStop}%
\bibitem [{\citenamefont {Einstein}\ \emph {et~al.}(1935)\citenamefont
  {Einstein}, \citenamefont {Podolsky},\ and\ \citenamefont {Rosen}}]{EPR_35}%
  \BibitemOpen
  \bibfield  {author} {\bibinfo {author} {\bibfnamefont {A.}~\bibnamefont
  {Einstein}}, \bibinfo {author} {\bibfnamefont {B.}~\bibnamefont {Podolsky}},
  \ and\ \bibinfo {author} {\bibfnamefont {N.}~\bibnamefont {Rosen}},\ }\href
  {\doibase 10.1103/PhysRev.47.777} {\bibfield  {journal} {\bibinfo  {journal}
  {Phys. Rev.}\ }\textbf {\bibinfo {volume} {47}},\ \bibinfo {pages} {777}
  (\bibinfo {year} {1935})}\BibitemShut {NoStop}%
\bibitem [{\citenamefont {Schrodinger}(1935)}]{S_35}%
  \BibitemOpen
  \bibfield  {author} {\bibinfo {author} {\bibfnamefont {E.}~\bibnamefont
  {Schrodinger}},\ }\href {\doibase 10.1017/S0305004100013554} {\bibfield
  {journal} {\bibinfo  {journal} {Mathematical Proceedings of the Cambridge
  Philosophical Society}\ }\textbf {\bibinfo {volume} {31}},\ \bibinfo {pages}
  {555} (\bibinfo {year} {1935})}\BibitemShut {NoStop}%
\bibitem [{\citenamefont {Reid}(1989)}]{R_89}%
  \BibitemOpen
  \bibfield  {author} {\bibinfo {author} {\bibfnamefont {M.~D.}\ \bibnamefont
  {Reid}},\ }\href {\doibase 10.1103/PhysRevA.40.913} {\bibfield  {journal}
  {\bibinfo  {journal} {Phys. Rev. A}\ }\textbf {\bibinfo {volume} {40}},\
  \bibinfo {pages} {913} (\bibinfo {year} {1989})}\BibitemShut {NoStop}%
\bibitem [{\citenamefont {Wiseman}\ \emph {et~al.}(2007)\citenamefont
  {Wiseman}, \citenamefont {Jones},\ and\ \citenamefont {Doherty}}]{WJD_07}%
  \BibitemOpen
  \bibfield  {author} {\bibinfo {author} {\bibfnamefont {H.~M.}\ \bibnamefont
  {Wiseman}}, \bibinfo {author} {\bibfnamefont {S.~J.}\ \bibnamefont {Jones}},
  \ and\ \bibinfo {author} {\bibfnamefont {A.~C.}\ \bibnamefont {Doherty}},\
  }\href {\doibase 10.1103/PhysRevLett.98.140402} {\bibfield  {journal}
  {\bibinfo  {journal} {Phys. Rev. Lett.}\ }\textbf {\bibinfo {volume} {98}},\
  \bibinfo {pages} {140402} (\bibinfo {year} {2007})}\BibitemShut {NoStop}%
\bibitem [{\citenamefont {Jones}\ \emph {et~al.}(2007)\citenamefont {Jones},
  \citenamefont {Wiseman},\ and\ \citenamefont {Doherty}}]{JWD_07}%
  \BibitemOpen
  \bibfield  {author} {\bibinfo {author} {\bibfnamefont {S.~J.}\ \bibnamefont
  {Jones}}, \bibinfo {author} {\bibfnamefont {H.~M.}\ \bibnamefont {Wiseman}},
  \ and\ \bibinfo {author} {\bibfnamefont {A.~C.}\ \bibnamefont {Doherty}},\
  }\href {\doibase 10.1103/PhysRevA.76.052116} {\bibfield  {journal} {\bibinfo
  {journal} {Phys. Rev. A}\ }\textbf {\bibinfo {volume} {76}},\ \bibinfo
  {pages} {052116} (\bibinfo {year} {2007})}\BibitemShut {NoStop}%
\bibitem [{\citenamefont {Skrzypczyk}\ \emph {et~al.}(2014)\citenamefont
  {Skrzypczyk}, \citenamefont {Navascu\'es},\ and\ \citenamefont
  {Cavalcanti}}]{SNC_14}%
  \BibitemOpen
  \bibfield  {author} {\bibinfo {author} {\bibfnamefont {P.}~\bibnamefont
  {Skrzypczyk}}, \bibinfo {author} {\bibfnamefont {M.}~\bibnamefont
  {Navascu\'es}}, \ and\ \bibinfo {author} {\bibfnamefont {D.}~\bibnamefont
  {Cavalcanti}},\ }\href {\doibase 10.1103/PhysRevLett.112.180404} {\bibfield
  {journal} {\bibinfo  {journal} {Phys. Rev. Lett.}\ }\textbf {\bibinfo
  {volume} {112}},\ \bibinfo {pages} {180404} (\bibinfo {year}
  {2014})}\BibitemShut {NoStop}%
\bibitem [{\citenamefont {Gallego}\ and\ \citenamefont {Aolita}(2015)}]{GA_15}%
  \BibitemOpen
  \bibfield  {author} {\bibinfo {author} {\bibfnamefont {R.}~\bibnamefont
  {Gallego}}\ and\ \bibinfo {author} {\bibfnamefont {L.}~\bibnamefont
  {Aolita}},\ }\href {\doibase 10.1103/PhysRevX.5.041008} {\bibfield  {journal}
  {\bibinfo  {journal} {Phys. Rev. X}\ }\textbf {\bibinfo {volume} {5}},\
  \bibinfo {pages} {041008} (\bibinfo {year} {2015})}\BibitemShut {NoStop}%
\bibitem [{\citenamefont {Bell}(1964)}]{B_64}%
  \BibitemOpen
  \bibfield  {author} {\bibinfo {author} {\bibfnamefont {J.~S.}\ \bibnamefont
  {Bell}},\ }\href {\doibase 10.1103/PhysicsPhysiqueFizika.1.195} {\bibfield
  {journal} {\bibinfo  {journal} {Physics Physique Fizika}\ }\textbf {\bibinfo
  {volume} {1}},\ \bibinfo {pages} {195} (\bibinfo {year} {1964})}\BibitemShut
  {NoStop}%
\bibitem [{\citenamefont {Clauser}\ \emph {et~al.}(1969)\citenamefont
  {Clauser}, \citenamefont {Horne}, \citenamefont {Shimony},\ and\
  \citenamefont {Holt}}]{CHSH_69}%
  \BibitemOpen
  \bibfield  {author} {\bibinfo {author} {\bibfnamefont {J.~F.}\ \bibnamefont
  {Clauser}}, \bibinfo {author} {\bibfnamefont {M.~A.}\ \bibnamefont {Horne}},
  \bibinfo {author} {\bibfnamefont {A.}~\bibnamefont {Shimony}}, \ and\
  \bibinfo {author} {\bibfnamefont {R.~A.}\ \bibnamefont {Holt}},\ }\href
  {\doibase 10.1103/PhysRevLett.23.880} {\bibfield  {journal} {\bibinfo
  {journal} {Phys. Rev. Lett.}\ }\textbf {\bibinfo {volume} {23}},\ \bibinfo
  {pages} {880} (\bibinfo {year} {1969})}\BibitemShut {NoStop}%
\bibitem [{\citenamefont {Werner}(1989)}]{W_89}%
  \BibitemOpen
  \bibfield  {author} {\bibinfo {author} {\bibfnamefont {R.~F.}\ \bibnamefont
  {Werner}},\ }\href {\doibase 10.1103/PhysRevA.40.4277} {\bibfield  {journal}
  {\bibinfo  {journal} {Phys. Rev. A}\ }\textbf {\bibinfo {volume} {40}},\
  \bibinfo {pages} {4277} (\bibinfo {year} {1989})}\BibitemShut {NoStop}%
\bibitem [{\citenamefont {Datta}\ \emph {et~al.}(2017)\citenamefont {Datta},
  \citenamefont {Goswami}, \citenamefont {Pramanik},\ and\ \citenamefont
  {Majumdar}}]{DGPM_17}%
  \BibitemOpen
  \bibfield  {author} {\bibinfo {author} {\bibfnamefont {S.}~\bibnamefont
  {Datta}}, \bibinfo {author} {\bibfnamefont {S.}~\bibnamefont {Goswami}},
  \bibinfo {author} {\bibfnamefont {T.}~\bibnamefont {Pramanik}}, \ and\
  \bibinfo {author} {\bibfnamefont {A.}~\bibnamefont {Majumdar}},\ }\href@noop
  {} {\bibfield  {journal} {\bibinfo  {journal} {Phys. Lett. A}\ }\textbf
  {\bibinfo {volume} {381}},\ \bibinfo {pages} {897} (\bibinfo {year}
  {2017})}\BibitemShut {NoStop}%
\bibitem [{\citenamefont {Aharonov}\ \emph {et~al.}(1988)\citenamefont
  {Aharonov}, \citenamefont {Albert},\ and\ \citenamefont {Vaidman}}]{AAV_88}%
  \BibitemOpen
  \bibfield  {author} {\bibinfo {author} {\bibfnamefont {Y.}~\bibnamefont
  {Aharonov}}, \bibinfo {author} {\bibfnamefont {D.~Z.}\ \bibnamefont
  {Albert}}, \ and\ \bibinfo {author} {\bibfnamefont {L.}~\bibnamefont
  {Vaidman}},\ }\href {\doibase 10.1103/PhysRevLett.60.1351} {\bibfield
  {journal} {\bibinfo  {journal} {Phys. Rev. Lett.}\ }\textbf {\bibinfo
  {volume} {60}},\ \bibinfo {pages} {1351} (\bibinfo {year}
  {1988})}\BibitemShut {NoStop}%
\bibitem [{\citenamefont {Dressel}\ \emph {et~al.}(2014)\citenamefont
  {Dressel}, \citenamefont {Malik}, \citenamefont {Miatto}, \citenamefont
  {Jordan},\ and\ \citenamefont {Boyd}}]{DMMJAB_14}%
  \BibitemOpen
  \bibfield  {author} {\bibinfo {author} {\bibfnamefont {J.}~\bibnamefont
  {Dressel}}, \bibinfo {author} {\bibfnamefont {M.}~\bibnamefont {Malik}},
  \bibinfo {author} {\bibfnamefont {F.~M.}\ \bibnamefont {Miatto}}, \bibinfo
  {author} {\bibfnamefont {A.~N.}\ \bibnamefont {Jordan}}, \ and\ \bibinfo
  {author} {\bibfnamefont {R.~W.}\ \bibnamefont {Boyd}},\ }\href {\doibase
  10.1103/RevModPhys.86.307} {\bibfield  {journal} {\bibinfo  {journal} {Rev.
  Mod. Phys.}\ }\textbf {\bibinfo {volume} {86}},\ \bibinfo {pages} {307}
  (\bibinfo {year} {2014})}\BibitemShut {NoStop}%
\bibitem [{\citenamefont {Hosten}\ and\ \citenamefont {Kwiat}(2008)}]{HK_08}%
  \BibitemOpen
  \bibfield  {author} {\bibinfo {author} {\bibfnamefont {O.}~\bibnamefont
  {Hosten}}\ and\ \bibinfo {author} {\bibfnamefont {P.}~\bibnamefont {Kwiat}},\
  }\href@noop {} {\bibfield  {journal} {\bibinfo  {journal} {Science}\ }\textbf
  {\bibinfo {volume} {319}},\ \bibinfo {pages} {787} (\bibinfo {year}
  {2008})}\BibitemShut {NoStop}%
\bibitem [{\citenamefont {Wiseman}(2002)}]{W_02}%
  \BibitemOpen
  \bibfield  {author} {\bibinfo {author} {\bibfnamefont {H.~M.}\ \bibnamefont
  {Wiseman}},\ }\href {\doibase 10.1103/PhysRevA.65.032111} {\bibfield
  {journal} {\bibinfo  {journal} {Phys. Rev. A}\ }\textbf {\bibinfo {volume}
  {65}},\ \bibinfo {pages} {032111} (\bibinfo {year} {2002})}\BibitemShut
  {NoStop}%
\bibitem [{\citenamefont {Solli}\ \emph {et~al.}(2004)\citenamefont {Solli},
  \citenamefont {McCormick}, \citenamefont {Chiao}, \citenamefont {Popescu},\
  and\ \citenamefont {Hickmann}}]{AMCPH_04}%
  \BibitemOpen
  \bibfield  {author} {\bibinfo {author} {\bibfnamefont {D.~R.}\ \bibnamefont
  {Solli}}, \bibinfo {author} {\bibfnamefont {C.~F.}\ \bibnamefont
  {McCormick}}, \bibinfo {author} {\bibfnamefont {R.~Y.}\ \bibnamefont
  {Chiao}}, \bibinfo {author} {\bibfnamefont {S.}~\bibnamefont {Popescu}}, \
  and\ \bibinfo {author} {\bibfnamefont {J.~M.}\ \bibnamefont {Hickmann}},\
  }\href {\doibase 10.1103/PhysRevLett.92.043601} {\bibfield  {journal}
  {\bibinfo  {journal} {Phys. Rev. Lett.}\ }\textbf {\bibinfo {volume} {92}},\
  \bibinfo {pages} {043601} (\bibinfo {year} {2004})}\BibitemShut {NoStop}%
\bibitem [{\citenamefont {Brunner}\ \emph {et~al.}(2004)\citenamefont
  {Brunner}, \citenamefont {Scarani}, \citenamefont {Wegm{\"u}ller},
  \citenamefont {Legr{\'e}},\ and\ \citenamefont {Gisin}}]{BSWLG_04}%
  \BibitemOpen
  \bibfield  {author} {\bibinfo {author} {\bibfnamefont {N.}~\bibnamefont
  {Brunner}}, \bibinfo {author} {\bibfnamefont {V.}~\bibnamefont {Scarani}},
  \bibinfo {author} {\bibfnamefont {M.}~\bibnamefont {Wegm{\"u}ller}}, \bibinfo
  {author} {\bibfnamefont {M.}~\bibnamefont {Legr{\'e}}}, \ and\ \bibinfo
  {author} {\bibfnamefont {N.}~\bibnamefont {Gisin}},\ }\href@noop {}
  {\bibfield  {journal} {\bibinfo  {journal} {Physical review letters}\
  }\textbf {\bibinfo {volume} {93}},\ \bibinfo {pages} {203902} (\bibinfo
  {year} {2004})}\BibitemShut {NoStop}%
\bibitem [{\citenamefont {Lundeen}\ \emph {et~al.}(2011)\citenamefont
  {Lundeen}, \citenamefont {Sutherland}, \citenamefont {Patel}, \citenamefont
  {Stewart},\ and\ \citenamefont {Bamber}}]{LSPSB_11}%
  \BibitemOpen
  \bibfield  {author} {\bibinfo {author} {\bibfnamefont {J.~S.}\ \bibnamefont
  {Lundeen}}, \bibinfo {author} {\bibfnamefont {B.}~\bibnamefont {Sutherland}},
  \bibinfo {author} {\bibfnamefont {A.}~\bibnamefont {Patel}}, \bibinfo
  {author} {\bibfnamefont {C.}~\bibnamefont {Stewart}}, \ and\ \bibinfo
  {author} {\bibfnamefont {C.}~\bibnamefont {Bamber}},\ }\href@noop {}
  {\bibfield  {journal} {\bibinfo  {journal} {Nature}\ }\textbf {\bibinfo
  {volume} {474}},\ \bibinfo {pages} {188} (\bibinfo {year}
  {2011})}\BibitemShut {NoStop}%
\bibitem [{\citenamefont {Brunner}\ and\ \citenamefont {Simon}(2010)}]{BS_10}%
  \BibitemOpen
  \bibfield  {author} {\bibinfo {author} {\bibfnamefont {N.}~\bibnamefont
  {Brunner}}\ and\ \bibinfo {author} {\bibfnamefont {C.}~\bibnamefont
  {Simon}},\ }\href {\doibase 10.1103/PhysRevLett.105.010405} {\bibfield
  {journal} {\bibinfo  {journal} {Phys. Rev. Lett.}\ }\textbf {\bibinfo
  {volume} {105}},\ \bibinfo {pages} {010405} (\bibinfo {year}
  {2010})}\BibitemShut {NoStop}%
\bibitem [{\citenamefont {Kocsis}\ \emph {et~al.}(2011)\citenamefont {Kocsis},
  \citenamefont {Braverman}, \citenamefont {Ravets}, \citenamefont {Stevens},
  \citenamefont {Mirin}, \citenamefont {Shalm},\ and\ \citenamefont
  {Steinberg}}]{KBRSMSS_11}%
  \BibitemOpen
  \bibfield  {author} {\bibinfo {author} {\bibfnamefont {S.}~\bibnamefont
  {Kocsis}}, \bibinfo {author} {\bibfnamefont {B.}~\bibnamefont {Braverman}},
  \bibinfo {author} {\bibfnamefont {S.}~\bibnamefont {Ravets}}, \bibinfo
  {author} {\bibfnamefont {M.~J.}\ \bibnamefont {Stevens}}, \bibinfo {author}
  {\bibfnamefont {R.~P.}\ \bibnamefont {Mirin}}, \bibinfo {author}
  {\bibfnamefont {L.~K.}\ \bibnamefont {Shalm}}, \ and\ \bibinfo {author}
  {\bibfnamefont {A.~M.}\ \bibnamefont {Steinberg}},\ }\href@noop {} {\bibfield
   {journal} {\bibinfo  {journal} {Science}\ }\textbf {\bibinfo {volume}
  {332}},\ \bibinfo {pages} {1170} (\bibinfo {year} {2011})}\BibitemShut
  {NoStop}%
\bibitem [{\citenamefont {Ghose}\ \emph {et~al.}(2001)\citenamefont {Ghose},
  \citenamefont {Majumdar}, \citenamefont {Guha},\ and\ \citenamefont
  {Sau}}]{GMGS_01}%
  \BibitemOpen
  \bibfield  {author} {\bibinfo {author} {\bibfnamefont {P.}~\bibnamefont
  {Ghose}}, \bibinfo {author} {\bibfnamefont {A.}~\bibnamefont {Majumdar}},
  \bibinfo {author} {\bibfnamefont {S.}~\bibnamefont {Guha}}, \ and\ \bibinfo
  {author} {\bibfnamefont {J.}~\bibnamefont {Sau}},\ }\href@noop {} {\bibfield
  {journal} {\bibinfo  {journal} {Physics Letters A}\ }\textbf {\bibinfo
  {volume} {290}},\ \bibinfo {pages} {205} (\bibinfo {year}
  {2001})}\BibitemShut {NoStop}%
\bibitem [{\citenamefont {Goswami}\ \emph {et~al.}(2019)\citenamefont
  {Goswami}, \citenamefont {Chakraborty}, \citenamefont {Ghosh},\ and\
  \citenamefont {Majumdar}}]{GCGM_18}%
  \BibitemOpen
  \bibfield  {author} {\bibinfo {author} {\bibfnamefont {S.}~\bibnamefont
  {Goswami}}, \bibinfo {author} {\bibfnamefont {S.}~\bibnamefont
  {Chakraborty}}, \bibinfo {author} {\bibfnamefont {S.}~\bibnamefont {Ghosh}},
  \ and\ \bibinfo {author} {\bibfnamefont {A.~S.}\ \bibnamefont {Majumdar}},\
  }\href {\doibase 10.1103/PhysRevA.99.012327} {\bibfield  {journal} {\bibinfo
  {journal} {Phys. Rev. A}\ }\textbf {\bibinfo {volume} {99}},\ \bibinfo
  {pages} {012327} (\bibinfo {year} {2019})}\BibitemShut {NoStop}%
\bibitem [{\citenamefont {Breuer}\ \emph {et~al.}(2002)\citenamefont {Breuer},
  \citenamefont {Petruccione} \emph {et~al.}}]{BP_02}%
  \BibitemOpen
  \bibfield  {author} {\bibinfo {author} {\bibfnamefont {H.-P.}\ \bibnamefont
  {Breuer}}, \bibinfo {author} {\bibfnamefont {F.}~\bibnamefont {Petruccione}},
   \emph {et~al.},\ }\href@noop {} {\emph {\bibinfo {title} {The theory of open
  quantum systems}}}\ (\bibinfo  {publisher} {Oxford University Press on
  Demand},\ \bibinfo {year} {2002})\BibitemShut {NoStop}%
\bibitem [{\citenamefont {Vedral}\ and\ \citenamefont {Plenio}(1998)}]{VP_98}%
  \BibitemOpen
  \bibfield  {author} {\bibinfo {author} {\bibfnamefont {V.}~\bibnamefont
  {Vedral}}\ and\ \bibinfo {author} {\bibfnamefont {M.~B.}\ \bibnamefont
  {Plenio}},\ }\href@noop {} {\bibfield  {journal} {\bibinfo  {journal}
  {Physical Review A}\ }\textbf {\bibinfo {volume} {57}},\ \bibinfo {pages}
  {1619} (\bibinfo {year} {1998})}\BibitemShut {NoStop}%
\bibitem [{\citenamefont {Buscemi}\ \emph
  {et~al.}(2003{\natexlab{a}})\citenamefont {Buscemi}, \citenamefont
  {D'Ariano}, \citenamefont {Perinotti},\ and\ \citenamefont
  {Sacchi}}]{BDPS_03}%
  \BibitemOpen
  \bibfield  {author} {\bibinfo {author} {\bibfnamefont {F.}~\bibnamefont
  {Buscemi}}, \bibinfo {author} {\bibfnamefont {G.}~\bibnamefont {D'Ariano}},
  \bibinfo {author} {\bibfnamefont {P.}~\bibnamefont {Perinotti}}, \ and\
  \bibinfo {author} {\bibfnamefont {M.}~\bibnamefont {Sacchi}},\ }\href@noop {}
  {\bibfield  {journal} {\bibinfo  {journal} {Physics Letters A}\ }\textbf
  {\bibinfo {volume} {314}},\ \bibinfo {pages} {374} (\bibinfo {year}
  {2003}{\natexlab{a}})}\BibitemShut {NoStop}%
\bibitem [{\citenamefont {Caruso}\ \emph {et~al.}(2011)\citenamefont {Caruso},
  \citenamefont {Eisert}, \citenamefont {Giovannetti},\ and\ \citenamefont
  {Holevo}}]{FEGVH_11}%
  \BibitemOpen
  \bibfield  {author} {\bibinfo {author} {\bibfnamefont {F.}~\bibnamefont
  {Caruso}}, \bibinfo {author} {\bibfnamefont {J.}~\bibnamefont {Eisert}},
  \bibinfo {author} {\bibfnamefont {V.}~\bibnamefont {Giovannetti}}, \ and\
  \bibinfo {author} {\bibfnamefont {A.~S.}\ \bibnamefont {Holevo}},\ }\href
  {\doibase 10.1103/PhysRevA.84.022306} {\bibfield  {journal} {\bibinfo
  {journal} {Phys. Rev. A}\ }\textbf {\bibinfo {volume} {84}},\ \bibinfo
  {pages} {022306} (\bibinfo {year} {2011})}\BibitemShut {NoStop}%
\bibitem [{\citenamefont {Buscemi}\ \emph
  {et~al.}(2003{\natexlab{b}})\citenamefont {Buscemi}, \citenamefont
  {D'Ariano},\ and\ \citenamefont {Sacchi}}]{BDS_18}%
  \BibitemOpen
  \bibfield  {author} {\bibinfo {author} {\bibfnamefont {F.}~\bibnamefont
  {Buscemi}}, \bibinfo {author} {\bibfnamefont {G.~M.}\ \bibnamefont
  {D'Ariano}}, \ and\ \bibinfo {author} {\bibfnamefont {M.~F.}\ \bibnamefont
  {Sacchi}},\ }\href {\doibase 10.1103/PhysRevA.68.042113} {\bibfield
  {journal} {\bibinfo  {journal} {Phys. Rev. A}\ }\textbf {\bibinfo {volume}
  {68}},\ \bibinfo {pages} {042113} (\bibinfo {year}
  {2003}{\natexlab{b}})}\BibitemShut {NoStop}%
\bibitem [{\citenamefont {Ziman}\ and\ \citenamefont {Bu\ifmmode~\check{z}\else
  \v{z}\fi{}ek}(2005)}]{ZB_05}%
  \BibitemOpen
  \bibfield  {author} {\bibinfo {author} {\bibfnamefont {M.}~\bibnamefont
  {Ziman}}\ and\ \bibinfo {author} {\bibfnamefont {V.}~\bibnamefont
  {Bu\ifmmode~\check{z}\else \v{z}\fi{}ek}},\ }\href {\doibase
  10.1103/PhysRevA.72.022343} {\bibfield  {journal} {\bibinfo  {journal} {Phys.
  Rev. A}\ }\textbf {\bibinfo {volume} {72}},\ \bibinfo {pages} {022343}
  (\bibinfo {year} {2005})}\BibitemShut {NoStop}%
\bibitem [{\citenamefont {G{\"u}hne}\ and\ \citenamefont
  {T{\'o}th}(2009)}]{GT_09}%
  \BibitemOpen
  \bibfield  {author} {\bibinfo {author} {\bibfnamefont {O.}~\bibnamefont
  {G{\"u}hne}}\ and\ \bibinfo {author} {\bibfnamefont {G.}~\bibnamefont
  {T{\'o}th}},\ }\href@noop {} {\bibfield  {journal} {\bibinfo  {journal}
  {Physics Reports}\ }\textbf {\bibinfo {volume} {474}},\ \bibinfo {pages} {1}
  (\bibinfo {year} {2009})}\BibitemShut {NoStop}%
\bibitem [{\citenamefont {Coffman}\ \emph {et~al.}(2000)\citenamefont
  {Coffman}, \citenamefont {Kundu},\ and\ \citenamefont {Wootters}}]{CKW_00}%
  \BibitemOpen
  \bibfield  {author} {\bibinfo {author} {\bibfnamefont {V.}~\bibnamefont
  {Coffman}}, \bibinfo {author} {\bibfnamefont {J.}~\bibnamefont {Kundu}}, \
  and\ \bibinfo {author} {\bibfnamefont {W.~K.}\ \bibnamefont {Wootters}},\
  }\href {\doibase 10.1103/PhysRevA.61.052306} {\bibfield  {journal} {\bibinfo
  {journal} {Phys. Rev. A}\ }\textbf {\bibinfo {volume} {61}},\ \bibinfo
  {pages} {052306} (\bibinfo {year} {2000})}\BibitemShut {NoStop}%
\bibitem [{\citenamefont {Cavalcanti}\ \emph {et~al.}(2015)\citenamefont
  {Cavalcanti}, \citenamefont {Foster}, \citenamefont {Fuwa},\ and\
  \citenamefont {Wiseman}}]{CFFW_15}%
  \BibitemOpen
  \bibfield  {author} {\bibinfo {author} {\bibfnamefont {E.~G.}\ \bibnamefont
  {Cavalcanti}}, \bibinfo {author} {\bibfnamefont {C.~J.}\ \bibnamefont
  {Foster}}, \bibinfo {author} {\bibfnamefont {M.}~\bibnamefont {Fuwa}}, \ and\
  \bibinfo {author} {\bibfnamefont {H.~M.}\ \bibnamefont {Wiseman}},\
  }\href@noop {} {\bibfield  {journal} {\bibinfo  {journal} {JOSA B}\ }\textbf
  {\bibinfo {volume} {32}},\ \bibinfo {pages} {A74} (\bibinfo {year}
  {2015})}\BibitemShut {NoStop}%
\bibitem [{not()}]{note}%
  \BibitemOpen
  \href@noop {} {}\bibinfo {note} {Note here that the reduced CPTP map (acting
  on $B$), formed out of $U_{SB}^{-1}$, is not necessarily the inverse of the
  given GADC (even if such an inverse map exists). This would have been the
  case, if under a suitable choice of an initial state $\sigma_{S}$ of the
  ancilla $S$, the reduced state $Tr_{S} [U_{SB}^{-1} (\sigma_{S} \otimes
  \Lambda(\rho_{B})) U_{SB}]$ of $B$ becomes close to the initial state
  $\rho_{B}$ before applying the GADC $\Lambda$.}\BibitemShut {Stop}%
\bibitem [{\citenamefont {Zanardi}\ \emph {et~al.}(2000)\citenamefont
  {Zanardi}, \citenamefont {Zalka},\ and\ \citenamefont {Faoro}}]{ZZF_00}%
  \BibitemOpen
  \bibfield  {author} {\bibinfo {author} {\bibfnamefont {P.}~\bibnamefont
  {Zanardi}}, \bibinfo {author} {\bibfnamefont {C.}~\bibnamefont {Zalka}}, \
  and\ \bibinfo {author} {\bibfnamefont {L.}~\bibnamefont {Faoro}},\ }\href
  {\doibase 10.1103/PhysRevA.62.030301} {\bibfield  {journal} {\bibinfo
  {journal} {Phys. Rev. A}\ }\textbf {\bibinfo {volume} {62}},\ \bibinfo
  {pages} {030301} (\bibinfo {year} {2000})}\BibitemShut {NoStop}%
\bibitem [{\citenamefont {Karimipour}\ \emph {et~al.}(2020)\citenamefont
  {Karimipour}, \citenamefont {Benatti},\ and\ \citenamefont
  {Floreanini}}]{KBF_20}%
  \BibitemOpen
  \bibfield  {author} {\bibinfo {author} {\bibfnamefont {V.}~\bibnamefont
  {Karimipour}}, \bibinfo {author} {\bibfnamefont {F.}~\bibnamefont {Benatti}},
  \ and\ \bibinfo {author} {\bibfnamefont {R.}~\bibnamefont {Floreanini}},\
  }\href {\doibase 10.1103/PhysRevA.101.032109} {\bibfield  {journal} {\bibinfo
   {journal} {Phys. Rev. A}\ }\textbf {\bibinfo {volume} {101}},\ \bibinfo
  {pages} {032109} (\bibinfo {year} {2020})}\BibitemShut {NoStop}%
\end{thebibliography}%

\end{document}